\documentclass{revtex4}
\usepackage{amsthm,amssymb,amsmath}				
\usepackage{graphicx,comment}
\usepackage{float}   
\usepackage{xcolor}

\begin{document}

\title{Confined Klein-Gordon oscillator from a (2+1)-dimensional G\"{u}rses to a G\"{u}rses or a pseudo-G\"{u}rses space-time backgrounds: Invariance and isospectrality}
\author{Omar Mustafa}
\email{omar.mustafa@emu.edu.tr}
\affiliation{Department of Physics, Eastern Mediterranean University, G. Magusa, north
Cyprus, Mersin 10 - Turkey.}

\begin{abstract}
\textbf{Abstract:}\ We study the Klein-Gordon (KG) oscillator with a Cornell-type scalar confinement in (2+1)-dimensional G\"{u}rses space-time backgrounds and report their exact solutions. The effect of the vorticity parameter $\Omega$ on the energy levels is found to yield some interesting features like; energy levels-crossings, partial clustering of positive and negative energy levels, and shifting the energy gap upwards or downwards. Such confined KG-oscillators are also studied in a general deformed G\"{u}rses space-time background. Moreover, we consider the confined-deformed KG-oscillator from a (2+1)-dimensional G\"{u}rses to G\"{u}rses and pseudo-G\"{u}rses space-time backgrounds. The resulting confined-deformed  KG-oscillators are found to admit invariance and isospectrality with each other.   


\end{abstract}

\maketitle

\section{Introduction}

Inspired by the Dirac oscillator \cite{Moshinsky 1989}, the Klein-Gordon
(KG) oscillator \cite{Bruce 1993,Dvoeg 1994} has been a subject of intensive
research in the last few decades. For example, the KG-oscillator in
the G\"{o}del and G\"{o}del-type space-time backgrounds (e.g., \cite%
{Moshinsky 1989,Bruce 1993,Dvoeg 1994,Das 2008,Carvalho 2016,Garcia
2017,Vitoria 2016,Vitoria1 2016,Ahmed1 2018}), in cosmic string space-time and
Kaluza-Klein theory backgrounds (e.g., \cite{Ahmed1 2018,Ahmed 2020,Ahmed1
2020,Ahmed1 2021,Boumal 2014}), in Som-Raychaudhuri \cite{Wang 2015}, in the
(2+1)-dimensional G\"{u}rses space-time backgrounds (e.g., \cite{Gurses
1994,Ahmed 2019,Ahmed1 2019,Ahmed2 2019}), etc. The reader may find a
sufficiently comprehensive sample of references on the historical progress
background of this issue in \cite{Ahmed 2020,Ahmed 2019,Lutfuoglu 2020,Mirza 2004,Deng 2019}. The KG-oscillator
in the (2+1)-dimensional G\"{u}rses space-time backgrounds is the focal
point of the current methodical proposal.

On the other hand, the introduction of Mathews-Lakshmanan oscillator \cite%
{M-L 1974} has activated intensive research studies on\emph{\ "effective"}
position-dependent mass (PDM in short), both in classical and quantum
mechanics 
{\cite{M-L 1974,von Roos,Carinena Ranada Sant 2004,Mustafa 2019,Mustafa 2020,Mustafa arXiv,
Mustafa Phys.Scr. 2020,Mustafa Habib 2007,Mustafa Algadhi 2019,Khlevniuk 2018,Mustafa 2015,
Dutra Almeida 2000,dos Santos 2021,Nabulsi1 2020,Nabulsi2 2020,Nabulsi3 2021,Quesne 2015,Tiwari 2013}}. PDM is a metaphoric manifestation of the coordinate deformation/transformation \cite%
{Mustafa 2020,Mustafa arXiv,Mustafa Phys.Scr. 2020}. Nevertheless, Khlevniuk 
\cite{Khlevniuk 2018} has argued that a point mass in the curved space may
effectively be transformed into a PDM in Euclidean space. Such coordinate
transformation/deformation affects, in turn, the form of the canonical momentum in
classical and the momentum operator in quantum mechanics (e.g., \cite%
{Mustafa 2020,Mustafa arXiv,Mustafa Algadhi 2019,dos Santos 2021} and
related references therein). In classical mechanics, it has been shown that
negative the gradient of the potential force field is no longer the time
derivative of the canonical momentum $p=m\left( x\right) \dot{x}$, but it is
rather related to the time derivative of the pseudo-momentum (also called
Noether momentum) $\pi \left( x\right) =\sqrt{m\left( x\right) }\dot{x}$ 
\cite{Mustafa arXiv}. In quantum mechanics, however, the PDM momentum
operator is constructed \cite{Mustafa Algadhi 2019} and used to find the PDM
creation and annihilation operators for the Schr\"{o}dinger oscillator 
\cite{Mustafa 2020}. It would be interesting, therefore, to investigate the
effects of such PDM recipe \cite{Mustafa 2020,Mustafa arXiv,Mustafa Algadhi
2019,dos Santos 2021} on the KG-oscillator in the (2+1)-dimensional G\"{u}rses space-time backgrounds with a confinement.

The KG-oscillator in a (1+2)-dimensional G\"{u}rses space-time backgrounds
was investigated by Ahmed \cite{Ahmed 2019,Ahmed1 2019}, without a
confinement (i.e., the scalar type interaction $S\left( r\right) =0$ in $%
m\longrightarrow m+S\left( r\right) $). In the current methodical proposal,
however, we consider the KG-oscillator confined in a Cornell-type scalar
potential (i.e., $S\left( r\right) $ of (\ref{Cornell potential}) below, which is commonly used in quarkonium  spectroscopy \cite{Quigg 1979,Lutfuoglu 2020}) in a (2+1)-dimensional G\"{u}rses space-time backgrounds. We discuss the confined KG-oscillator in a G\"{u}rses space-time background (i.e. G\"{u}rses space-time metric $ds^{2}$ (\ref{Gurses metric}) at specific G\"{u}rses parametric settings) and report the corresponding exact solution in section 2. Therein, we discuss and report the effects of the vorticity parameter (i.e., $\Omega$ in (\ref{Gurses metric}) below) on the energies levels through the reported Figures 1-4. Such figures exhibit some interesting features like, energy levels crossings, partial clustering of positive and negative energy levels, and shifting the energy  gap upwards or downwards. In section 3, we consider the confined KG-oscillator in a generalized \emph{deformed} (2+1)-dimensional G\"{u}rses space-time background. We show that
the resulting  confined-deformed KG-oscillator is, in fact, invariant and isospectral with that in the (2+1)-dimensional G\"{u}rses space-time background of section 1. We discuss, in section 4, a confined-deformed KG-oscillator from a (2+1)-dimensional G\"{u}rses to yet another G\"{u}rses
space-time backgrounds. That is, the deformation in the (2+1)-dimensional G\"{u}rses space-time metric $d\tilde{s}^{2}$ is chosen so that it belongs to a another G\"{u}rses space-time metric but with different  G\"{u}rses-type parametric settings. Moreover, we consider (in section 5) a confined-deformed KG-oscillator from G\"{u}rses to a pseudo-G\"{u}rses space-time backgrounds. The notion of \emph{pseudo-G\"{u}rses space-time metric} is manifested by the fact that its parametric settings do not belong to the set of parameters of G\"{u}rses space-time metric, but it can be transformed into G\"{u}rses space-time metric within a transformation (\ref{PT1}) below. The resulting  confined-deformed KG-oscillators are found to admit invariance and isospectrality with the confined KG-oscillator in a (2+1)-dimensional G\"{u}rses space-time background discussed in section 2. Our concluding remarks are given in section 6.

\section{Confined Klein-Gordon oscillator in a (2+1)-dimensional G\"{u}rses space-time background}

In this section, we recollect the basic formulation of the KG-oscillator in a (2+1)-dimensional G\"{u}rses space-time background. Hence,
we consider the (2+1)-dimensional G\"{u}rses space-time metric \cite{Gurses 1994}%
\begin{equation}
ds^{2}=-dt^{2}+dr^{2}-2\Omega r^{2}dtd\theta +r^{2}\left( 1-\Omega
^{2}r^{2}\right) d\theta ^{2}=g_{\mu \nu }dx^{\mu }dx^{\nu };\text{ }\mu
,\nu =0,1,2  \label{Gurses metric}
\end{equation}%
with $a_{_{0}}=b_{_{0}}=e_{_{0}}=1$, $b_{_{1}}=c_{_{0}}=%
\lambda _{_{0}}=0$, and the vorticity $\Omega =-\mu /3$, in the G\"{u}rses metric%
\begin{equation}
ds^{2}=-\phi dt^{2}+2qdtd\theta +\frac{h^{2}\psi -q^{2}}{a_{_{0}}}d\theta
^{2}+\frac{1}{\psi }dr^{2}  \label{Gurses metric0}
\end{equation}%
(i.e., as in Eq.(5) of \cite{Gurses 1994}), where%
\begin{equation}
\phi =a_{_{0}},\,\psi =b_{_{0}}+\frac{b_{_{1}}}{r^{2}}+\frac{3\lambda _{_{0}}%
}{4}r^{2},\,q=c_{_{0}}+\frac{e_{_{0}}\mu }{3}r^{2},\,h=e_{_{0}}r,\,\lambda
_{_{0}}=\lambda +\frac{\mu ^{2}}{27}.  \label{Gurses metric parameters}
\end{equation}%
The covariant and contravarian metric tensors in this case, respectively,
read%
\begin{equation}
g_{\mu \nu }=\left( 
\begin{tabular}{ccc}
$-1\smallskip $ & $0$ & $-\Omega r^{2}$ \\ 
$0$ & $1\smallskip $ & $0$ \\ 
$-\Omega r^{2}$ & $\,0$ & $\,r^{2}\left( 1-\Omega ^{2}r^{2}\right) $%
\end{tabular}%
\right) \Longleftrightarrow g^{\mu \nu }=\left( 
\begin{tabular}{ccc}
$-\left( 1-\Omega ^{2}r^{2}\right) $ & $0\smallskip $ & $-\Omega $ \\ 
$0$ & $1\smallskip $ & $0$ \\ 
$-\Omega $ & $\,0$ & $\,\frac{1}{r^{2}}$%
\end{tabular}%
\right) \text{ };\text{ \ }\det \left( g\right) =-r^{2}.  \label{g-tensors}
\end{equation}%
Under such setting, the KG-equation, with a scalar
confinement $S\left( r\right) $ (i.e., $m\longrightarrow m+S\left( r\right) $), is given by%
\begin{equation}
\frac{1}{\sqrt{-g}}\partial _{\mu }\left( \sqrt{-g}g^{\mu \nu }\partial
_{\nu }\Psi \right) =\left( m+S\left( r\right) \right) ^{2}\Psi .
\label{KG-eq}
\end{equation}%
Moreover, we may now couple the KG-oscillator using the recipes in \cite{Mirza 2004,Deng 2019} and allow the momentum operator to indulge the oscillator through%
\begin{equation}
p_{\mu }\longrightarrow p_{\mu }+i\eta \chi _{\mu },  \label{3D-momentum}
\end{equation}%
with $\eta $ denoting the frequency of the oscillator and $\chi _{\mu
}=\left( 0,r,0\right) $. This would, in effect, transform KG-equation (\ref{KG-eq}) into%
\begin{equation}
\frac{1}{\sqrt{-g}}\left( \partial _{\mu }+\eta \chi _{\mu }\right) \left[
\sqrt{-g}g^{\mu \nu }\left( \partial _{\nu }{\color{red}-}\eta\chi _{\nu }\right)
\Psi \right] =\left( m+S\left( r\right) \right) ^{2}\Psi .
\label{KG-oscillator}
\end{equation}%
Which consequently yields%
\begin{equation}
\left\{ -\partial _{t}^{2}+\left( \Omega \,r\,\partial _{t}-\frac{1}{r}%
\partial _{\theta }\right) ^{2}+\partial _{r}^{2}+\frac{1}{r}\partial
_{r}-\eta ^{2}r^{2}-2\eta -\left( m+S\left( r\right) \right)
^{2}\right\} \Psi =0.  \label{KG-O-1}
\end{equation}%
A textbook substitution in the form of%
\begin{equation}
\Psi \left( t,r,\theta \right) =\exp \left( i\left[ \ell \theta -Et\right]
\right) \psi \left( r\right) =\exp \left({\color{red}+} i\left[ \ell \theta -Et\right]
\right) \frac{R\left( r\right) }{\sqrt{r}}  \label{Psi(t,r,phi)}
\end{equation}%
would result in a one-dimensional Schr\"{o}dinger-like KG-oscillator with a confinement $S(r)$ so that %
\begin{equation}
R^{\prime \prime }\left( r\right) +\left[ \lambda -\frac{\left( \ell
^{2}-1/4\right) }{r^{2}}-\tilde{\omega}^{2}r^{2}-2mS\left( r\right) -S\left(
r\right) ^{2}\right] R\left( r\right) =0,  \label{R(r)-eq}
\end{equation}%
where%
\begin{equation}
\lambda =E^{2}-2\,\Omega \,\ell \,E-2\eta -m^{2}\text{ ; \ }\tilde{\omega}%
^{2}=\Omega ^{2}E^{2}+\eta ^{2}.  \label{lambda-omega}
\end{equation}%
Obviously, equation (\ref{R(r)-eq}) represents, with $S(r)=0$, the 2-dimensional radial harmonic oscillator with an effective oscillation frequency $\tilde{\omega}$ and consequently inherits its textbook eigenvalues%
\begin{equation}
\lambda =2 \tilde{\omega}\left( 2n_{r}+\left\vert \ell \right\vert +1\right)
\label{HO-lambda}
\end{equation}%
and radial eigenfunctions%
\begin{equation}
R\left( r\right) \sim r^{\left\vert \ell \right\vert +1/2}\exp \left( -\frac{%
\tilde{\omega}r^{2}}{2}\right) L_{n_{r}}^{\left\vert \ell \right\vert
}\left( \tilde{\omega}r^{2}\right) \Longleftrightarrow \psi \left( r\right)
\sim r^{\left\vert \ell \right\vert }\exp \left( -\frac{\tilde{\omega}r^{2}}{%
2}\right) L_{n_{r}}^{\left\vert \ell \right\vert }\left( \tilde{\omega}%
r^{2}\right) .  \label{HO-R(r)}
\end{equation}%
At this point we may move further and use a Cornell-type scalar potential%
\begin{equation}
S\left( r\right) =Ar+\frac{B}{r}  \label{Cornell potential}
\end{equation}%
so that equation (\ref{R(r)-eq}) now reads%
\begin{equation}
R^{\prime \prime }\left( r\right) +\left[ \tilde{\lambda}-\frac{\left( 
\tilde{\ell}^{2}-1/4\right) }{r^{2}}-\beta ^{2}r^{2}-2mAr-\frac{2mB}{r}%
\right] R\left( r\right) =0,  \label{R(r)-eq-S(r)}
\end{equation}%
where%
\begin{equation}
\tilde{\lambda}=E^{2}-2\,\Omega \,\ell \,E-2\eta -m^{2}-2AB,\text{ \ }%
\tilde{\ell}^{2}=\ell ^{2}+B^{2},\text{ }\beta ^{2}=\Omega
^{2}E^{2}+\eta ^{2}+A^{2}.  \label{parameters-S(r)}
\end{equation}%
Equation (\ref{R(r)-eq-S(r)}) admits a finite/bounded solution in the form
of biconfluent Heun functions %
\begin{equation}
R\left( r\right) \sim \mathcal{\,}r^{\left\vert \tilde{\ell}\right\vert
+1/2}\,\exp \left( -\frac{\beta ^{2}r^{2}+2Amr}{2\,\beta }\right)
\,H_{B}\left( 2\left\vert \tilde{\ell}\right\vert ,\frac{2mA}{\beta ^{3/2}},%
\frac{A^{2}m^{2}+\tilde{\lambda}\,\beta ^{2}}{\beta ^{3}},\frac{4mB}{\sqrt{%
\beta }},\sqrt{\beta }r\right),  \label{S(r)-solution}
\end{equation}%
where for  $A=B=0$, $\alpha'=2|\tilde\ell|=2|\ell|$ is a non-negative integer, $\beta'=\frac{2mA}{\beta ^{3/2}}=0$, $\gamma'=\frac{A^{2}m^{2}+\tilde{\lambda}\,\beta ^{2}}{\beta ^{3}}=\frac{\tilde{\lambda}}{\beta}$,  and $\delta'=\frac{4mB}{\sqrt{\beta }}=0$, in $H_{B}\left( \alpha' ,\beta' ,\gamma' ,\delta' ,z\right)$. Then the Kummer relation \cite{Ron 1995} in this case suggests that%
\begin{equation}
H_{B}\left( \alpha' ,0 ,\gamma' ,0 ,z\right)={_{1}F_{1}}\left( (\frac{1}{2}+\frac{\alpha'}{4}-\frac{\gamma'}{4}) ,(1+\frac{\alpha'}{2}),z^2\right)={_{1}F_{1}}\left( -n_r ,(1+\frac{\alpha'}{2}),z^2\right),
\label{Heun to hypergeo}
\end{equation}%
where $n_r=0,1,2,\cdots $ is the radial quantum number (number of nodes in the wave function). This would in turn establish the relation between $\alpha'$ and $\gamma'$ and allow us to write%
\begin{equation}
\frac{1}{2}+\frac{\alpha'}{4}-\frac{\gamma'}{4}=-n_r\Longleftrightarrow \gamma'=4n_r+\alpha'+2\Longleftrightarrow \alpha'=\gamma'-2(2n_r+1),
\label{gamma-alpha relation}
\end{equation}%
where $2n_r\geq 0\Longrightarrow n_r\geq 0$ satisfies the condition of a finite polynomial nature of the biconfluent Heun functions. Therefore, for our $H_{B}\left( \alpha' ,\beta' ,\gamma' ,\delta' ,z\right)$ in (\ref{S(r)-solution}) to imply a quantum mechanically viable and acceptable solution, we adopt the parametric relation (\ref{gamma-alpha relation}). This would consequently yield that%
\begin{equation}
\frac{A^{2}m^{2}+\tilde{\lambda}\,\beta ^{2}}{\beta ^{3}}=2\left(
2n_{r}+\left\vert \tilde{\ell}\right\vert +1\right) \Longleftrightarrow 
\tilde{\lambda}=2\beta \left( 2n_{r}+\left\vert \tilde{\ell}\right\vert
+1\right) -\frac{m^{2}A^{2}}{\beta ^{2}}.  \label{lambda-S(r)}
\end{equation}%
In this case, we get the relation for the energy eigenvalues as%
\begin{equation}
E^{2}-2\,\Omega \,\ell \,E-2\eta-m^{2}-2AB+\frac{m^{2}A^{2}}{\left( \Omega ^{2}E^{2}+\eta ^{2}+A^{2}\right) } 
=2\left( \Omega ^{2}E^{2}+\eta^{2}+A^{2}\right) ^{1/2}\left(
2n_{r}+\left\vert \tilde{\ell}\right\vert +1\right) .  \label{energy-eq}
\end{equation}%
One should notice that this result collapses into that in (\ref{HO-lambda}) when the parameters $A$ and $B$ of the Cornell potential (\ref{Cornell potential}) are set zeros. This is indeed the natural tendency of a viable solution of a more general case (i.e., with the Cornell confinement). Moreover, it is obvious that this equation is hard to solve analytically, not impossible though. Yet, it is clear that the second term $(-2\,\Omega \,\ell \,E)$ on the L.H.S. plays a critical role in shaping the fate of energy levels' $E_{n_r,\ell}$ structure when plotted against the vorticity parameter $\Omega$ for some different parametric values, and against the oscillator frequency $\omega$. The effect of the term $(-2\,\Omega \,\ell \,E_{\pm})$ is identified ($\pm$ stands for positive and negative  energy regions, respectively ) as follows: (i) the positive energy $E_+$, with $\Omega$ and $\ell$ are both positive or both negative, will be boosted upward for $\ell\neq0$, whereas (ii) the negative energy  $E_-$, with positive/negative $\Omega$ and negative/positive $\ell$, the energy levels are boosted downward for $\ell\neq0$. This is documented in Figures 1-2. In Figure 3, we observe that the effect of the second term is clearly exemplified through shifting the energy gap upwards (when $\Omega$ and $\ell$ are both positive or both negative) and downwards (when $\Omega$ is negative/positive and $\ell$ is positive/negative, respectively). As a result, we  observe that the energy levels crossings  (documented in Fig.1), the energy levels partial clustering (documented in Fig.2(b)-2(f)), and the energy gap shifting (documented in Fig.3(a) and 3(b)) are unavoidable manifestations in the process. Furthermore, when $\ell=0$ the second term effect dies out and the spectrum retains its regular ordering format, but remains infected with the effect of the vorticity parameter though.%
\begin{figure}[!ht]  
\centering
\includegraphics[width=0.3\textwidth]{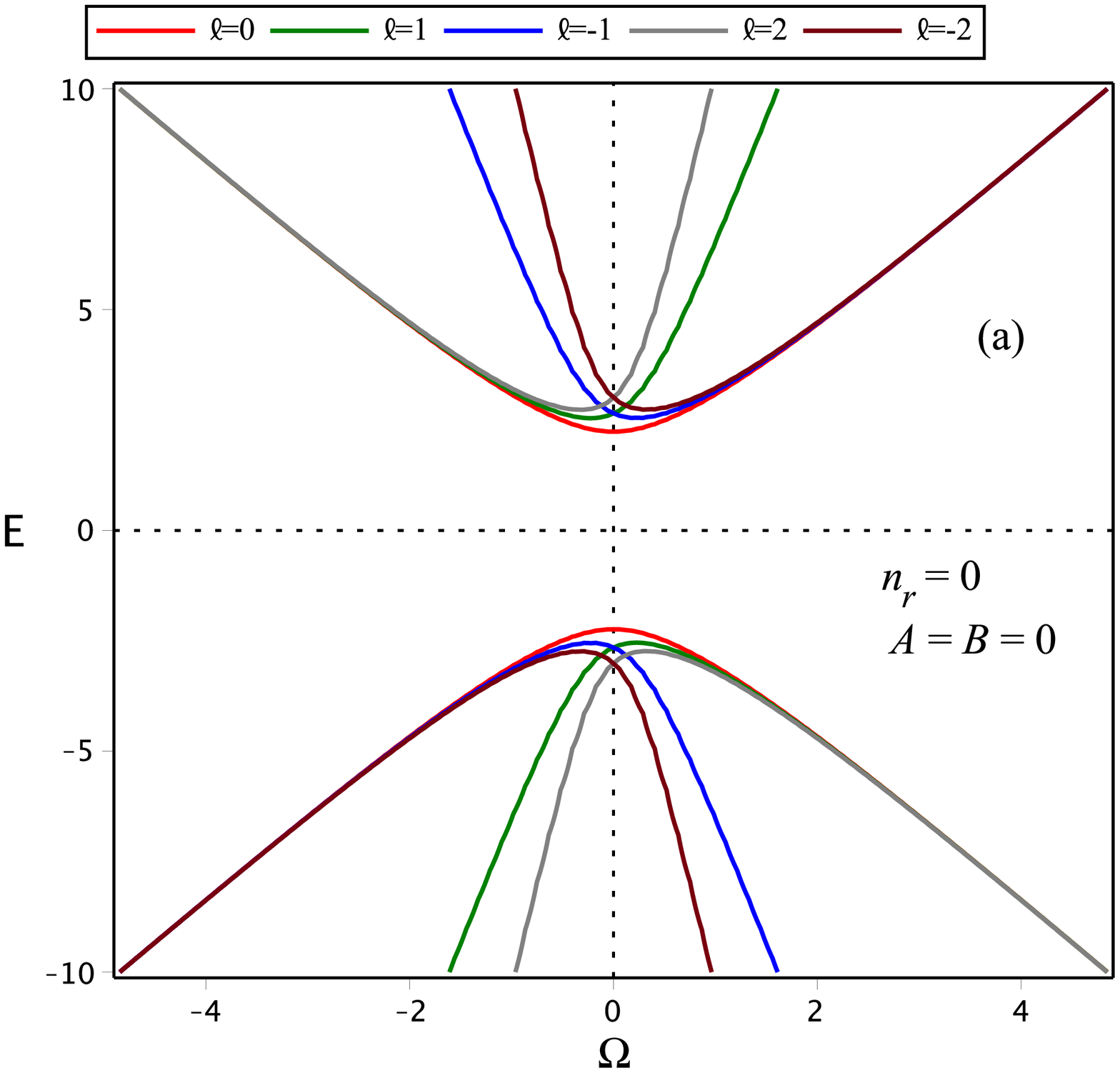}
\includegraphics[width=0.3\textwidth]{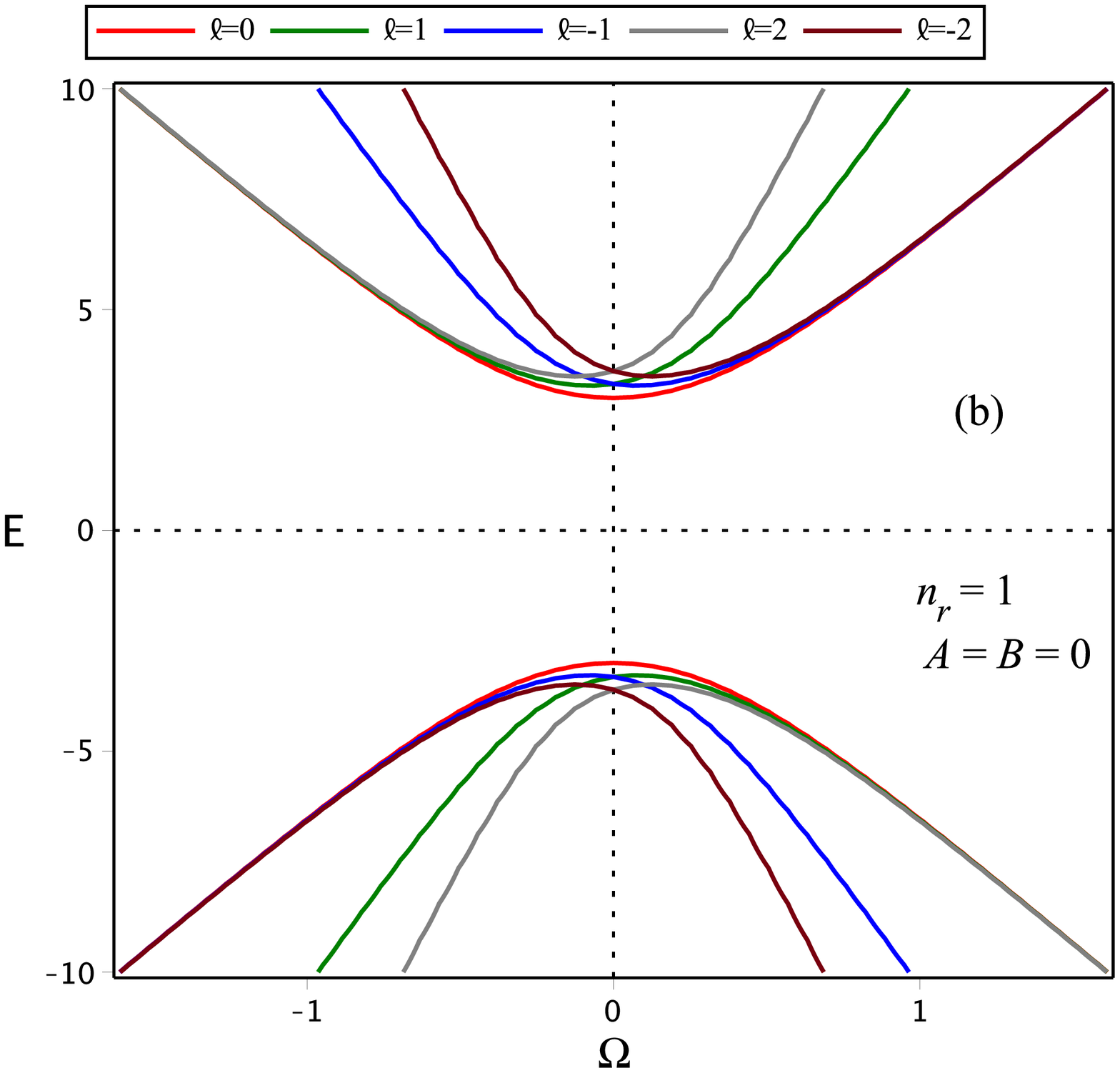}
\includegraphics[width=0.3\textwidth]{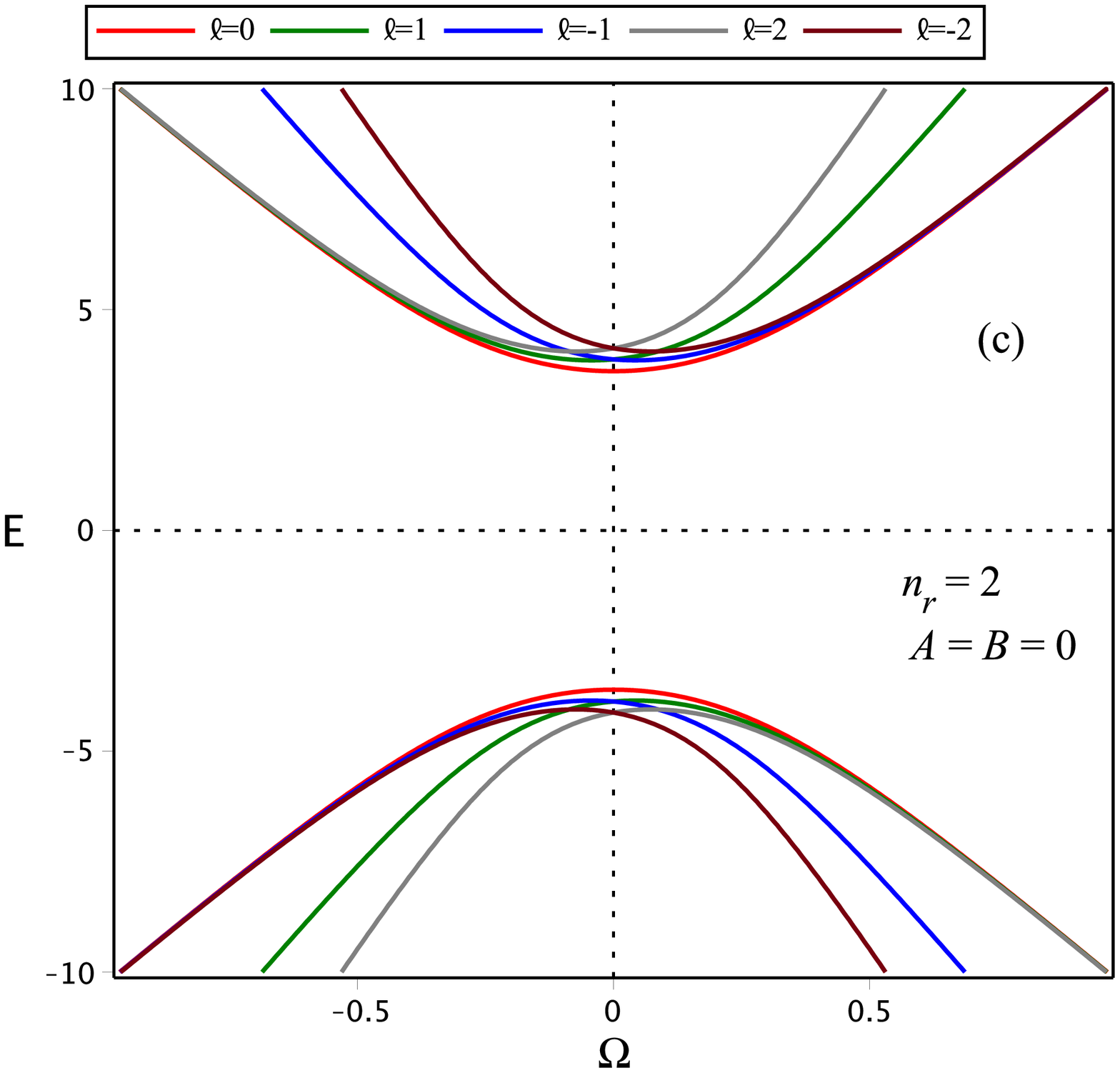}
\includegraphics[width=0.3\textwidth]{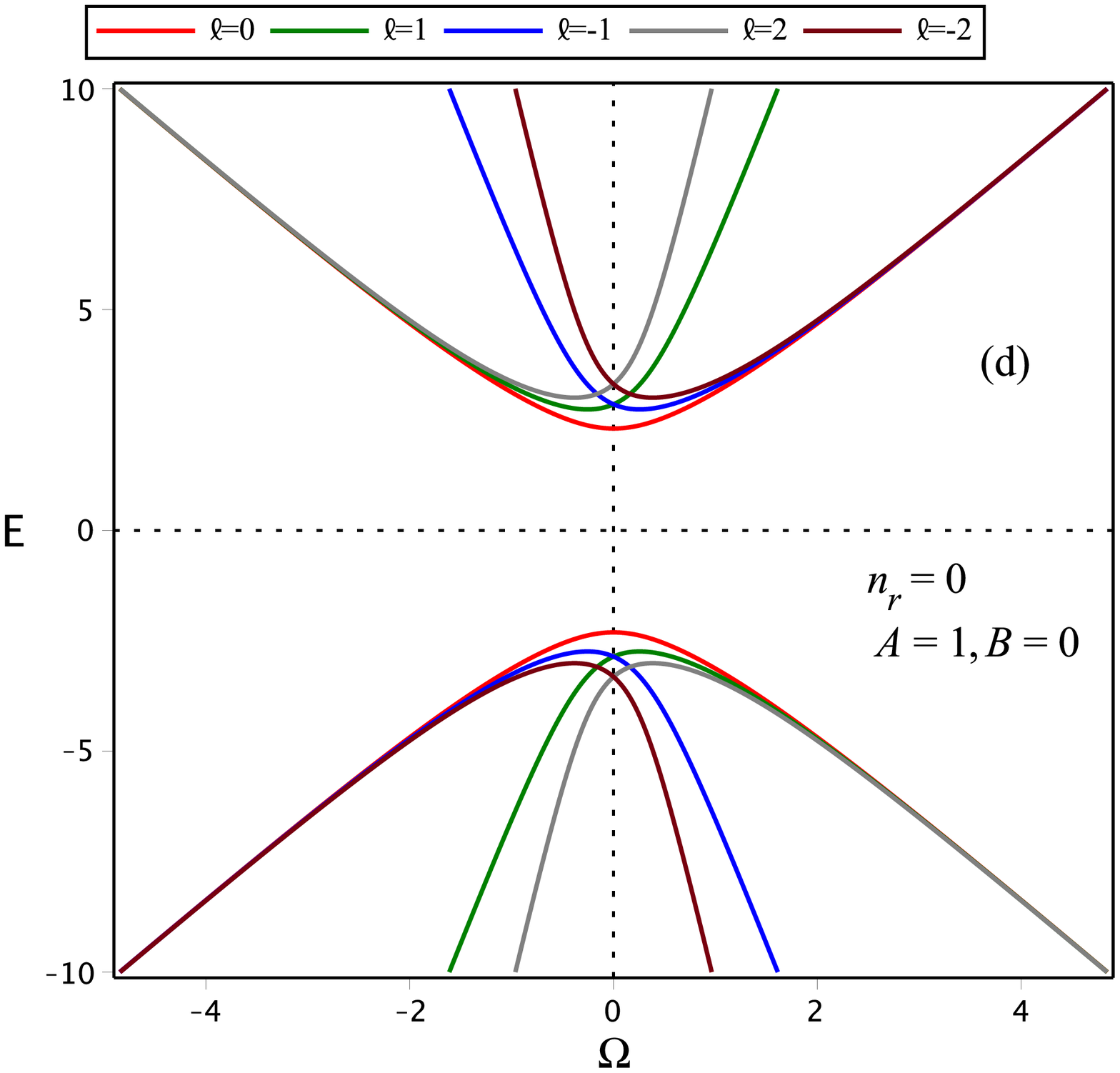}
\includegraphics[width=0.3\textwidth]{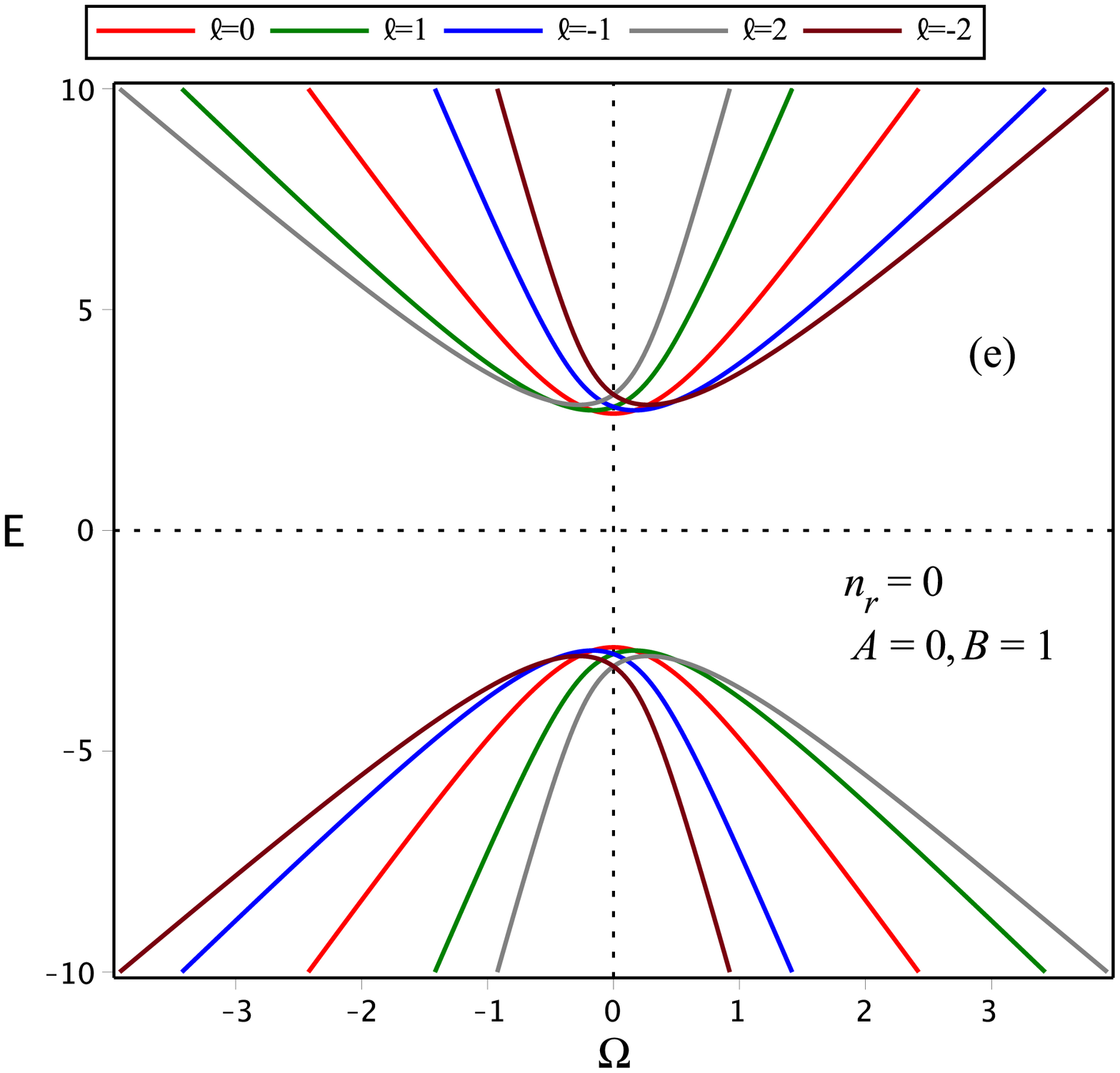}
\includegraphics[width=0.3\textwidth]{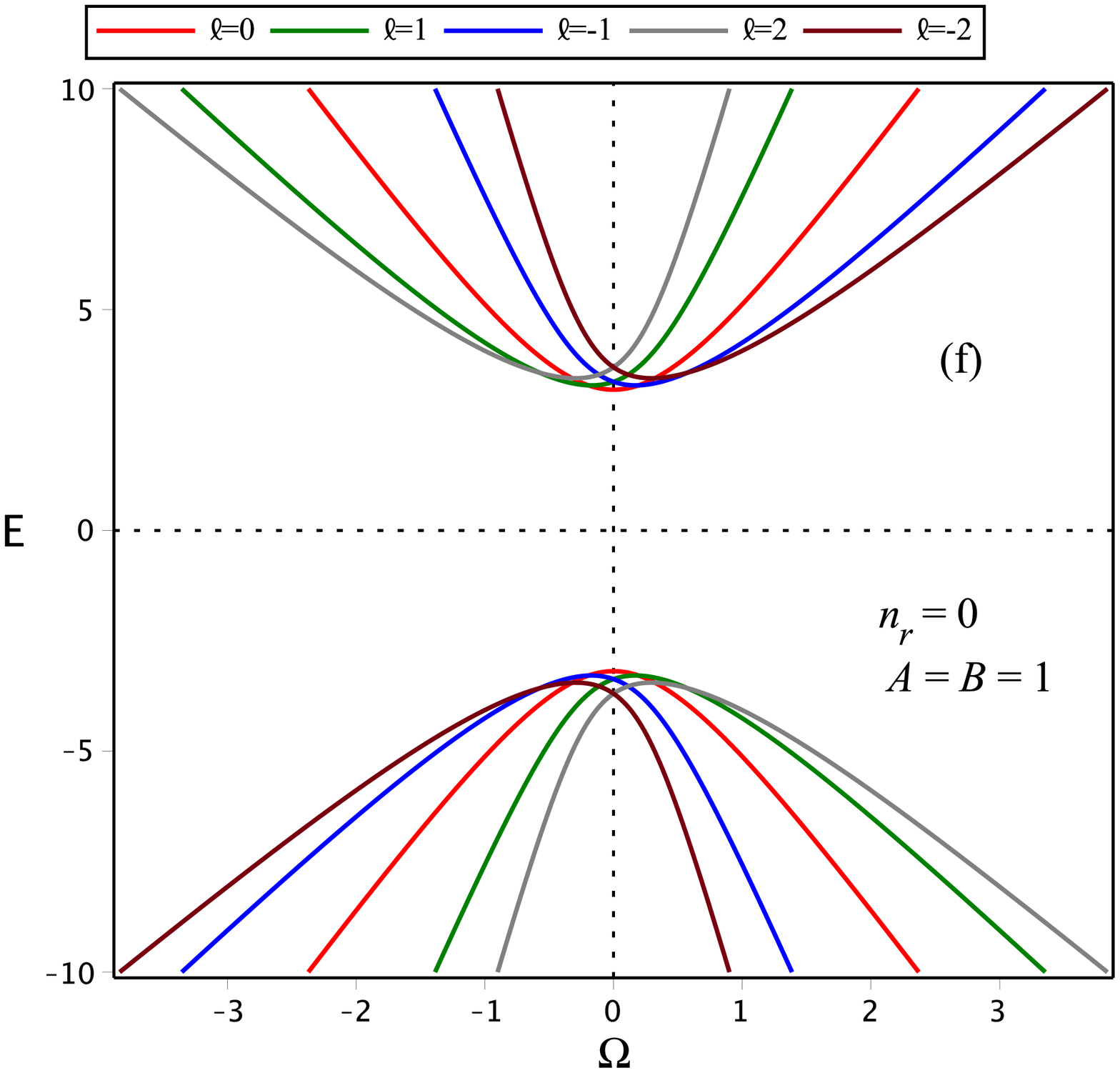}
\caption{\small 
{ For $\eta=m=1$ we show $E_{n_r,\ell}$ of (\ref{energy-eq}) against the vorticity parameter $\Omega$ and for $\ell=0,\pm1,\pm2$. In (a), (b), and (c) we show the KG-oscillator energies without the confinement (i.e., $A=B=0$) for $n_r=0,1,2$, respectively. In (d), (e), and (f) we show the parametric effects of the Cornell  confinement  (\ref{Cornell potential}) on the KG-oscillator energies with $n_r=0$ for $(A=1, B=0)$,  $(A=0, B=1)$, and  $(A=1,B=1)$, respectively.}}
\label{fig1}
\end{figure}%
\begin{figure}[!ht]  
\centering
\includegraphics[width=0.3\textwidth]{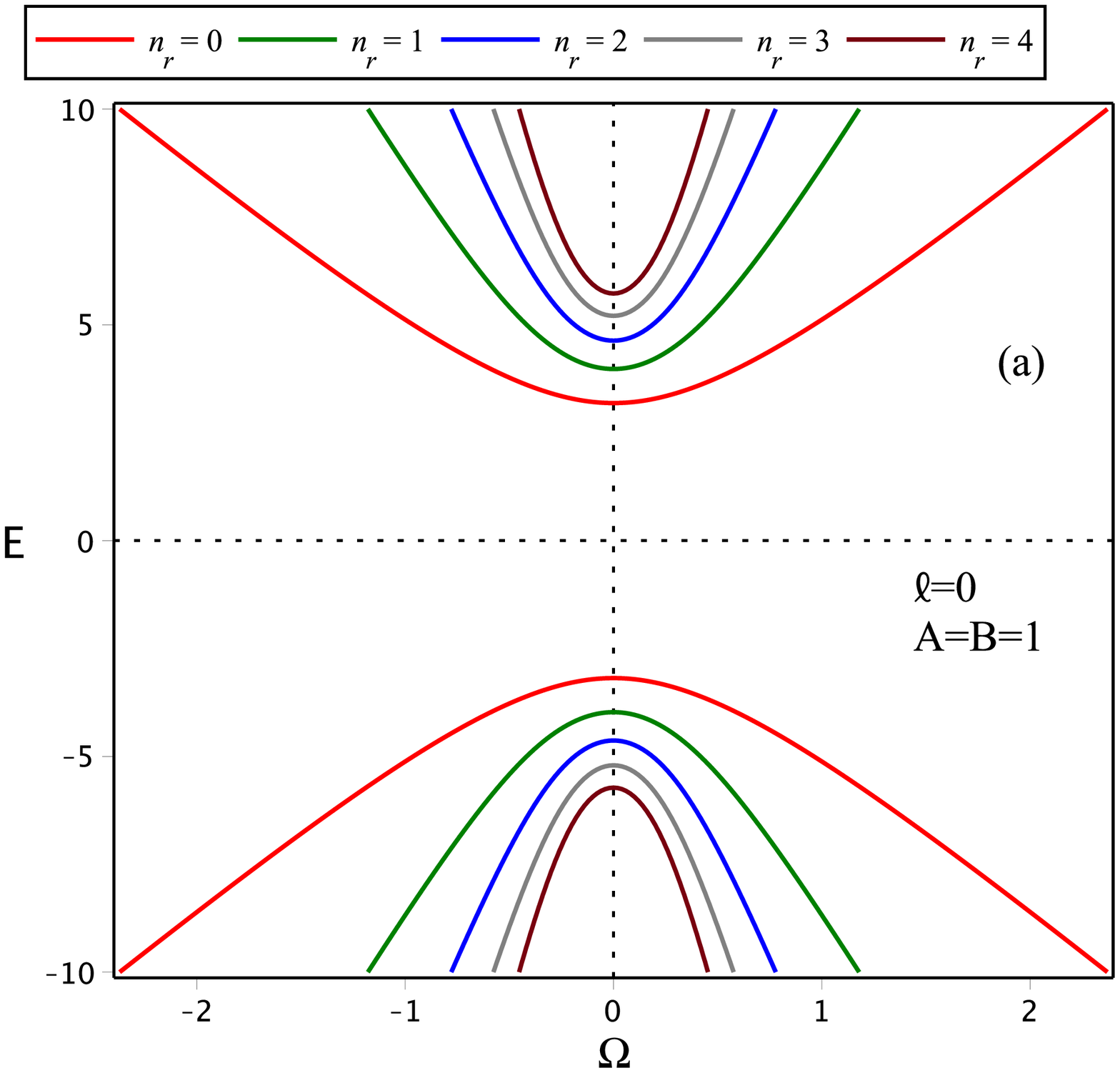}
\includegraphics[width=0.3\textwidth]{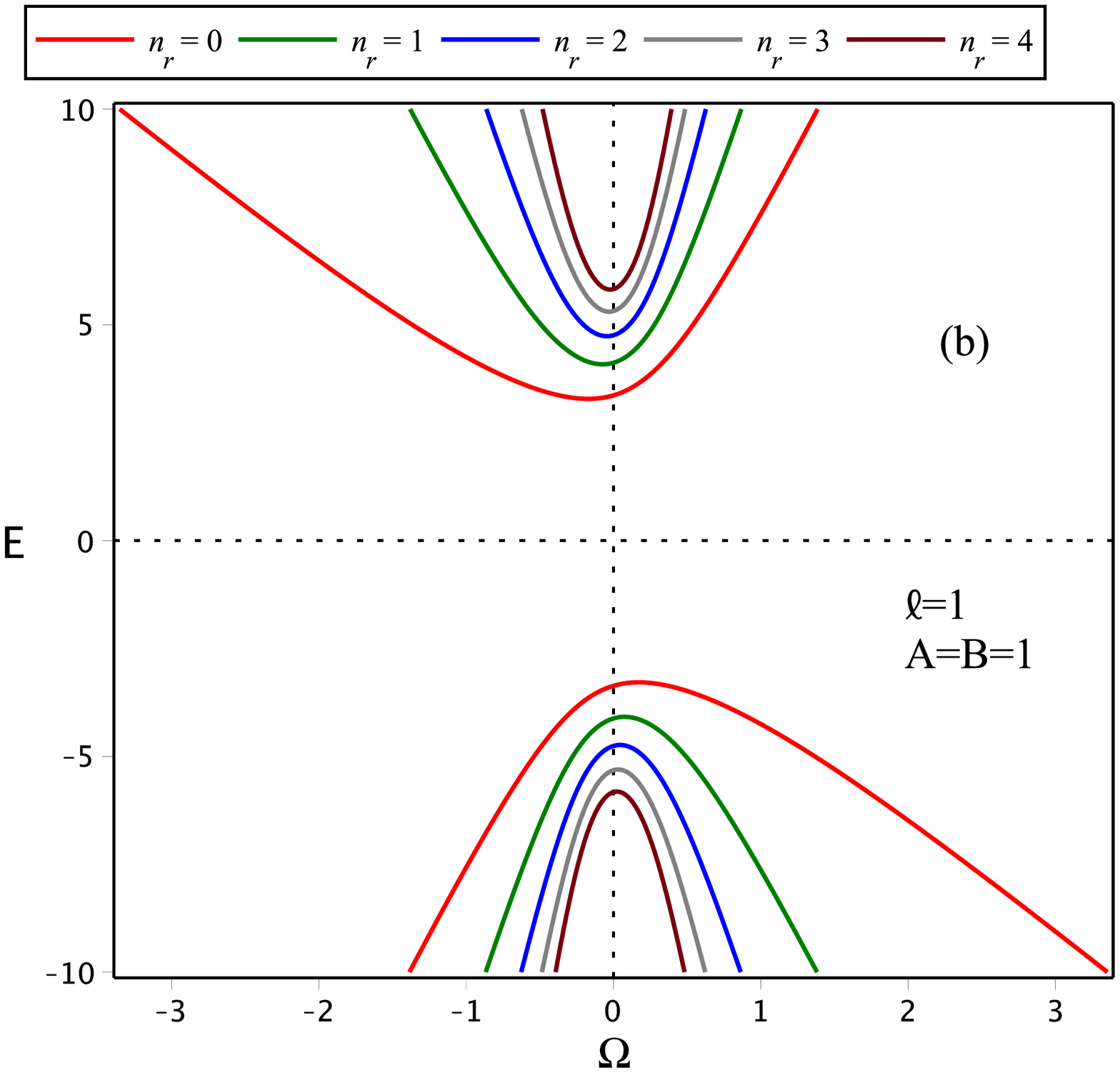}
\includegraphics[width=0.3\textwidth]{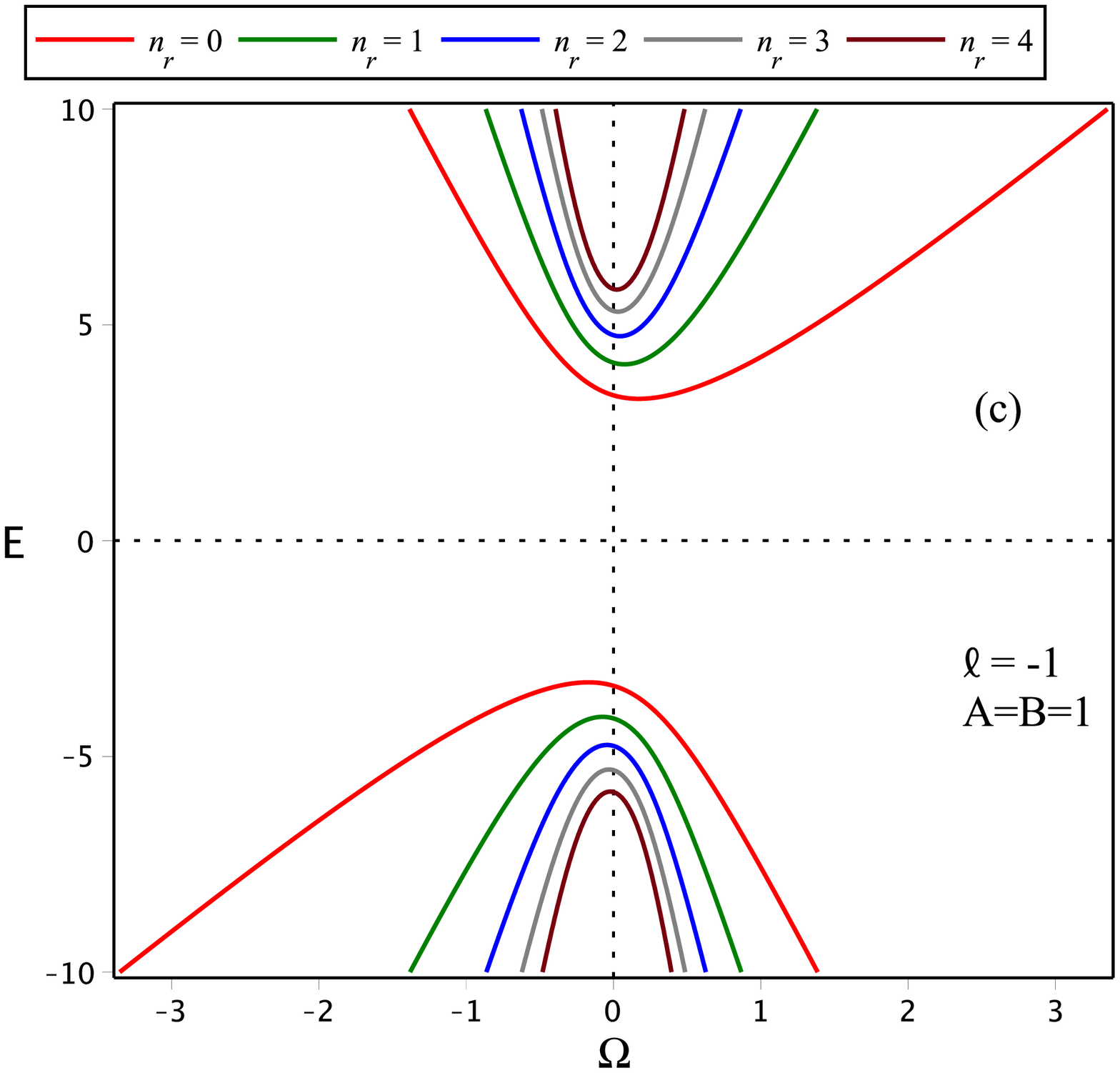}
\includegraphics[width=0.3\textwidth]{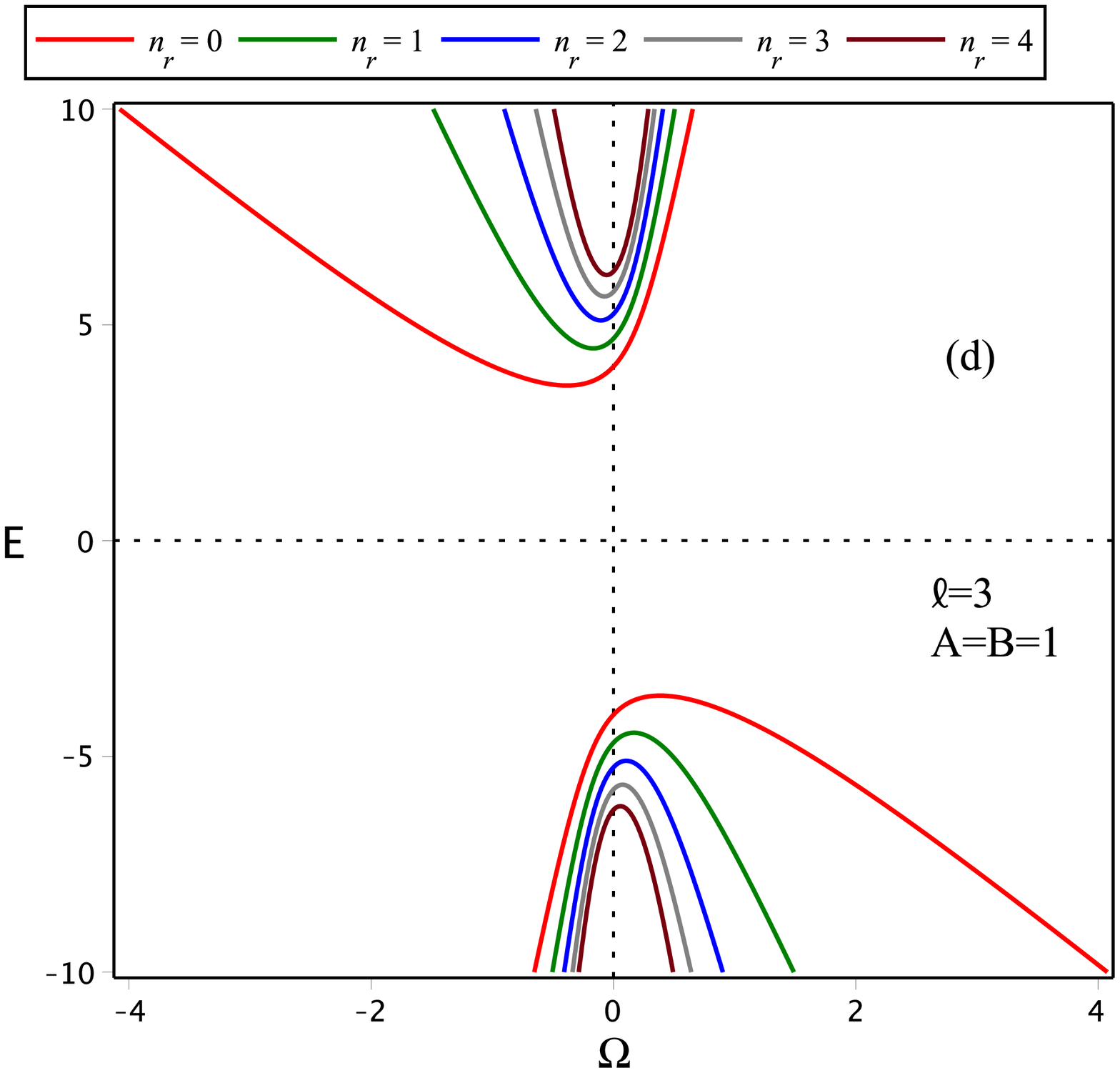}
\includegraphics[width=0.3\textwidth]{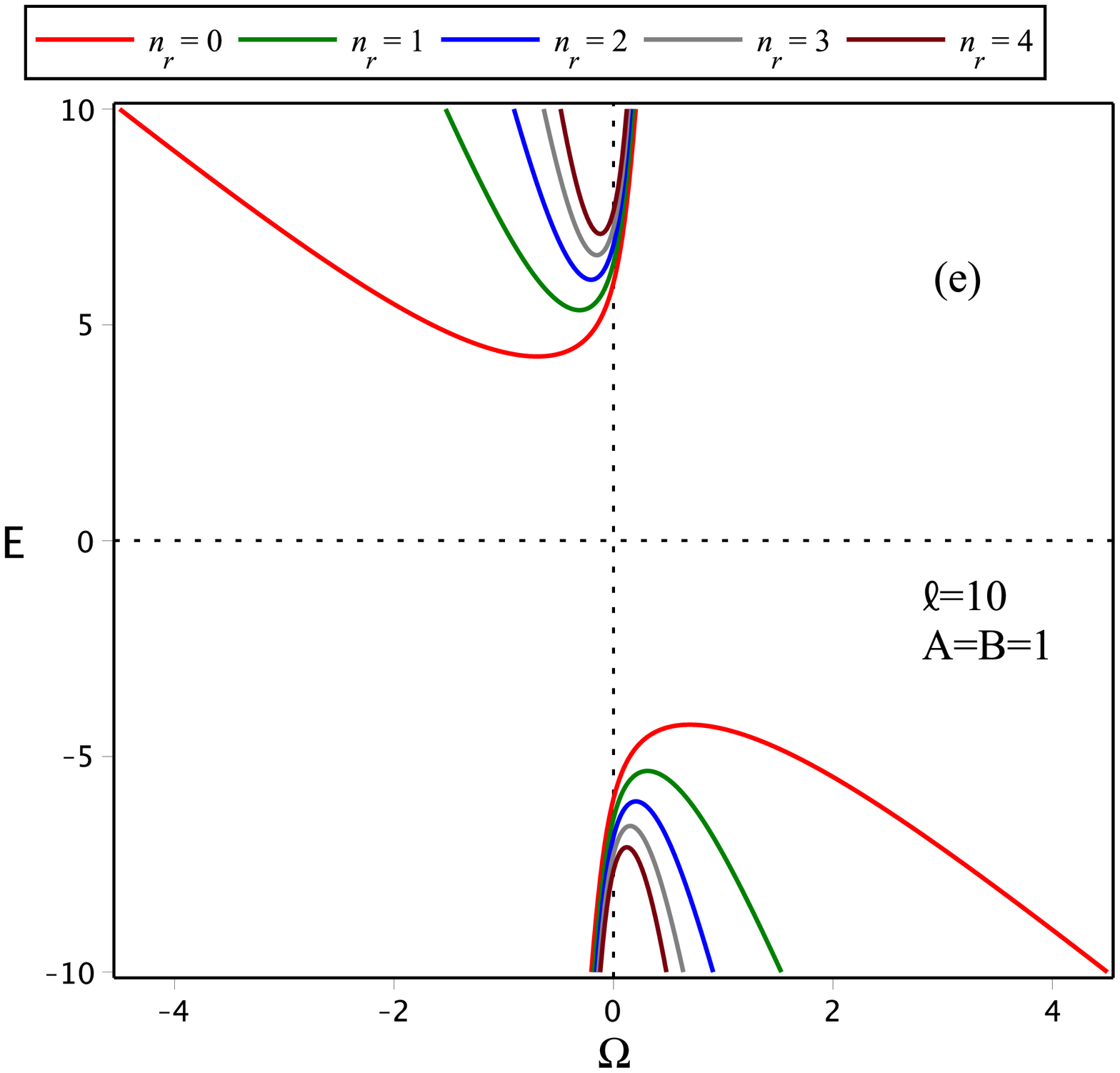}
\includegraphics[width=0.3\textwidth]{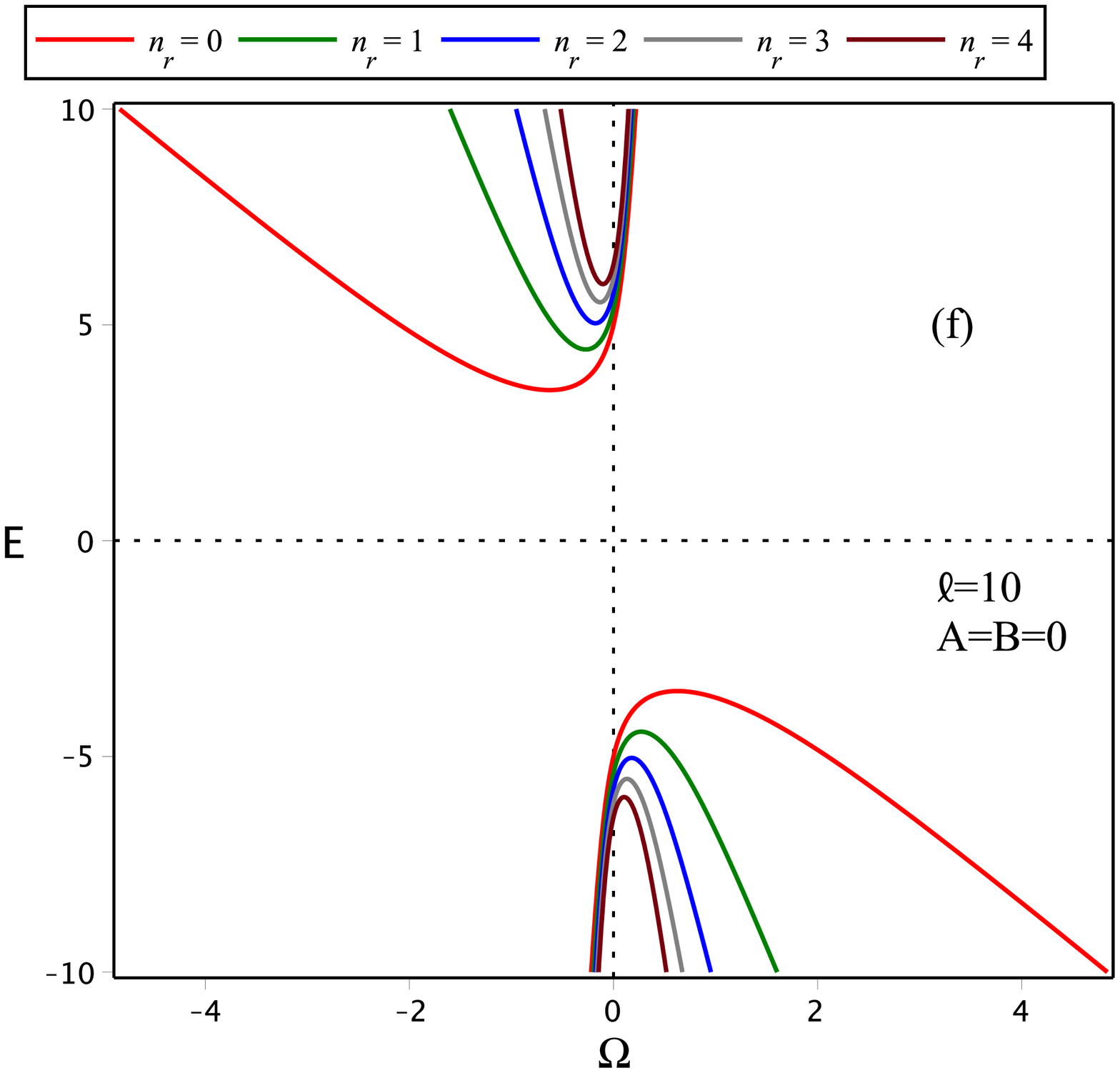}
\caption{\small 
{ For $\eta=m=1$ we show $E_{n_r,\ell}$ of (\ref{energy-eq}) against the vorticity parameter $\Omega$ at $n_r=0,1,2,3,4$. In (a), (b), (c), (d), and (e) we show the KG-oscillator energies with the Cornell confinement (\ref{Cornell potential}) (i.e., $A=B=1$) for $\ell=0,1,-1,3,10$, respectively. In (f) we show the KG-oscillator energies without the Cornell  confinement  (\ref{Cornell potential}) (i.e., $A=B=0$) at $\ell=10$.}}
\label{fig2}
\end{figure}%

\section{Confined KG-oscillator in a deformed (2+1)-dimensional G\"{u}rses space-time background}

In this section, we consider that the (2+1)-dimensional G\"{u}rses space-time
metric (\ref{Gurses metric}) is deformed in such a way that%
\begin{equation}
ds^{2}\longrightarrow d\tilde{s}^{2}=-\left( dt+\Omega \,\tilde{r}^{2}d%
\tilde{\theta}\right) ^{2}+\tilde{r}^{2}d\tilde{\theta}^{2}+d\tilde{r}^{2}.
\label{deformed Gurses metric}
\end{equation}%
Then the corresponding confined KG-oscillator is given, with $\tilde{\chi}%
_{\mu }=\left( 0,\tilde{r},0\right) $, by%
\begin{equation}
\frac{1}{\sqrt{-\tilde{g}}}\left( \,\tilde{\partial}_{\mu }+\eta \,\tilde{%
\chi}_{\mu }\right) \left[ \sqrt{-\tilde{g}}\tilde{g}^{\mu \nu }\left( \,%
\tilde{\partial}_{\nu }-\eta \,\tilde{\chi}_{\nu }\right) \Psi \left( t,%
\tilde{r},\tilde{\theta}\right) \right] =\left( m+A\tilde{r}+\frac{B}{\tilde{%
r}}\right) ^{2}\Psi \left( t,\tilde{r},\tilde{\theta}\right) .
\label{PDM-KG1}
\end{equation}%
We may now use the transformation%
\begin{equation}
\tilde{r}=\int \sqrt{m\left( r\right) }dr=\sqrt{Q\left( r\right) }r\text{ ;
\ }d\tilde{\theta}=d\theta \text{.}  \label{PT1}
\end{equation}%
to connect the deformed G\"{u}rses space-time (\ref{deformed Gurses metric})
with the formal one in (\ref{Gurses metric}). Moreover, the relation between 
$m\left( r\right) $ and $Q\left( r\right) $ is given by%
\begin{equation}
\frac{d\tilde{r}}{dr}=\sqrt{m\left( r\right) }=\sqrt{Q\left( r\right) }%
\left( 1+\frac{Q^{\prime }\left( r\right) }{2Q\left( r\right) }r\right)
\label{m(r)-Q(r)}
\end{equation}%
In this case, the deformed G\"{u}rses metric (\ref{deformed Gurses metric}) reads%
\begin{equation}
d\tilde{s}^{2}=-\left( dt+\Omega \,Q\left( r\right) r^{2}d\theta \right)
^{2}+Q\left( r\right) r^{2}d\theta ^{2}+m\left( r\right) dr^{2},
\label{Gurses transformed metric}
\end{equation}%
and consequently the corresponding space-time metric tensor is%
\begin{equation}
\tilde{g}_{\mu \nu }=\left( 
\begin{tabular}{ccc}
$-1\,$ & $0$ & $-\Omega \,Q\left( r\right) r^{2}\smallskip $ \\ 
$0$ & $\,m\left( r\right) \smallskip $ & $0$ \\ 
$-\Omega \,Q\left( r\right) r^{2}\smallskip $ & $0$ & $\,\,Q\left( r\right)
r^{2}\left( 1-\Omega ^{2}Q\left( r\right) r^{2}\right) $%
\end{tabular}%
\right) ,  \label{gij}
\end{equation}%
with its determinant $\det \left( \tilde{g}\right) =-m\left( r\right)
Q\left( r\right) r^{2}$ and its inverse%
\begin{equation}
\tilde{g}^{\mu \nu }=\left( 
\begin{tabular}{ccc}
$-\left( 1-\Omega ^{2}Q\left( r\right) r^{2}\right) \smallskip $ & $0$ & $%
-\Omega \,$ \\ 
$0\smallskip $ & $\frac{1}{m\left( r\right) }\,$ & $0$ \\ 
$-\Omega \,$ & $0$ & $\,\,\frac{1}{Q\left( r\right) r^{2}}$%
\end{tabular}%
\right) .  \label{inverse g}
\end{equation}%
The deformed (2+1)-dimensional space-time metric (\ref{Gurses transformed metric}) may very well be called a pseudo-G\"{u}rses metric for it may return to the G\"{u}rses one through reversing the transformation recipe.%
\begin{figure}[!ht]
\centering
\includegraphics[width=0.3\textwidth]{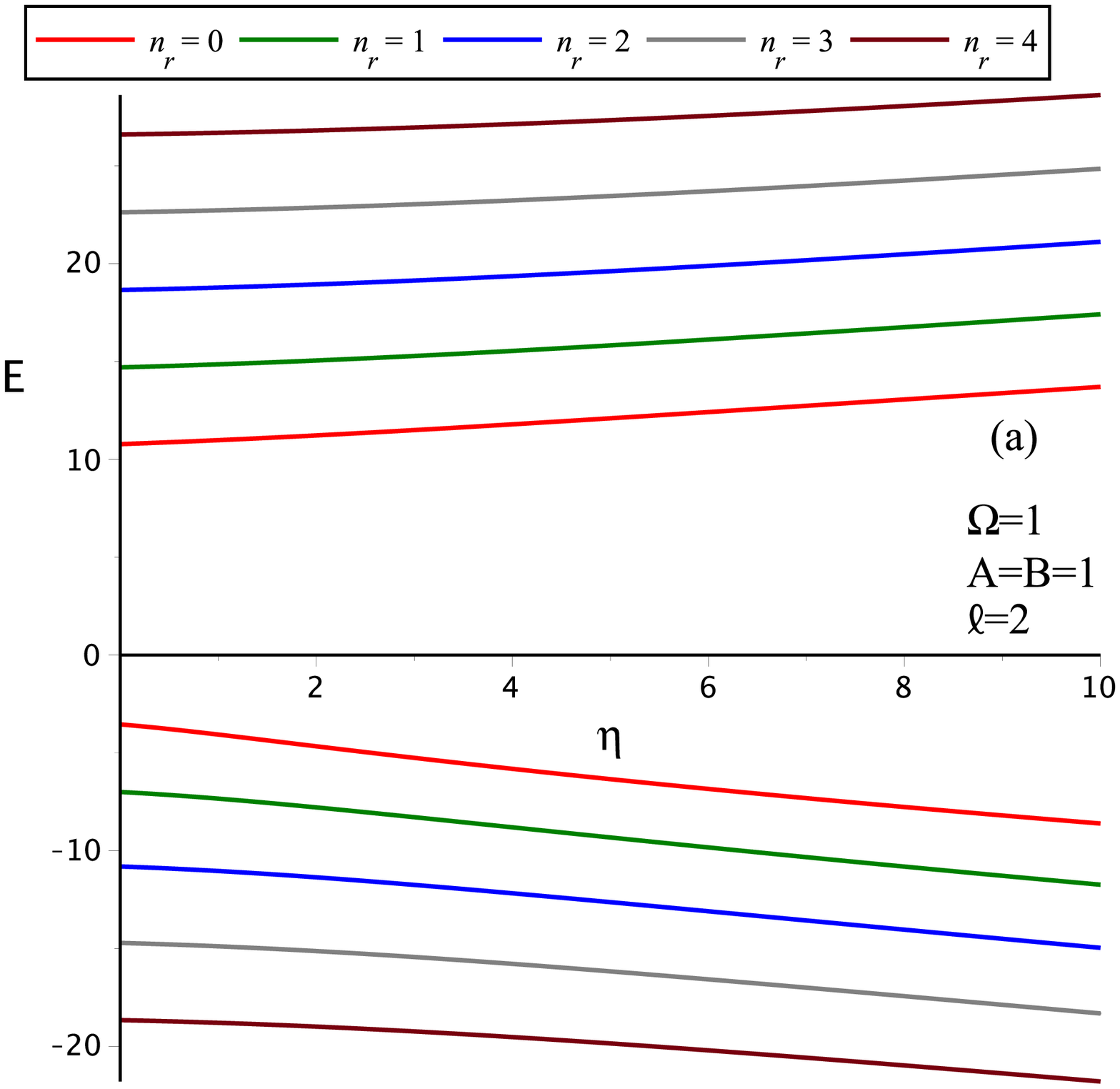}
\includegraphics[width=0.3\textwidth]{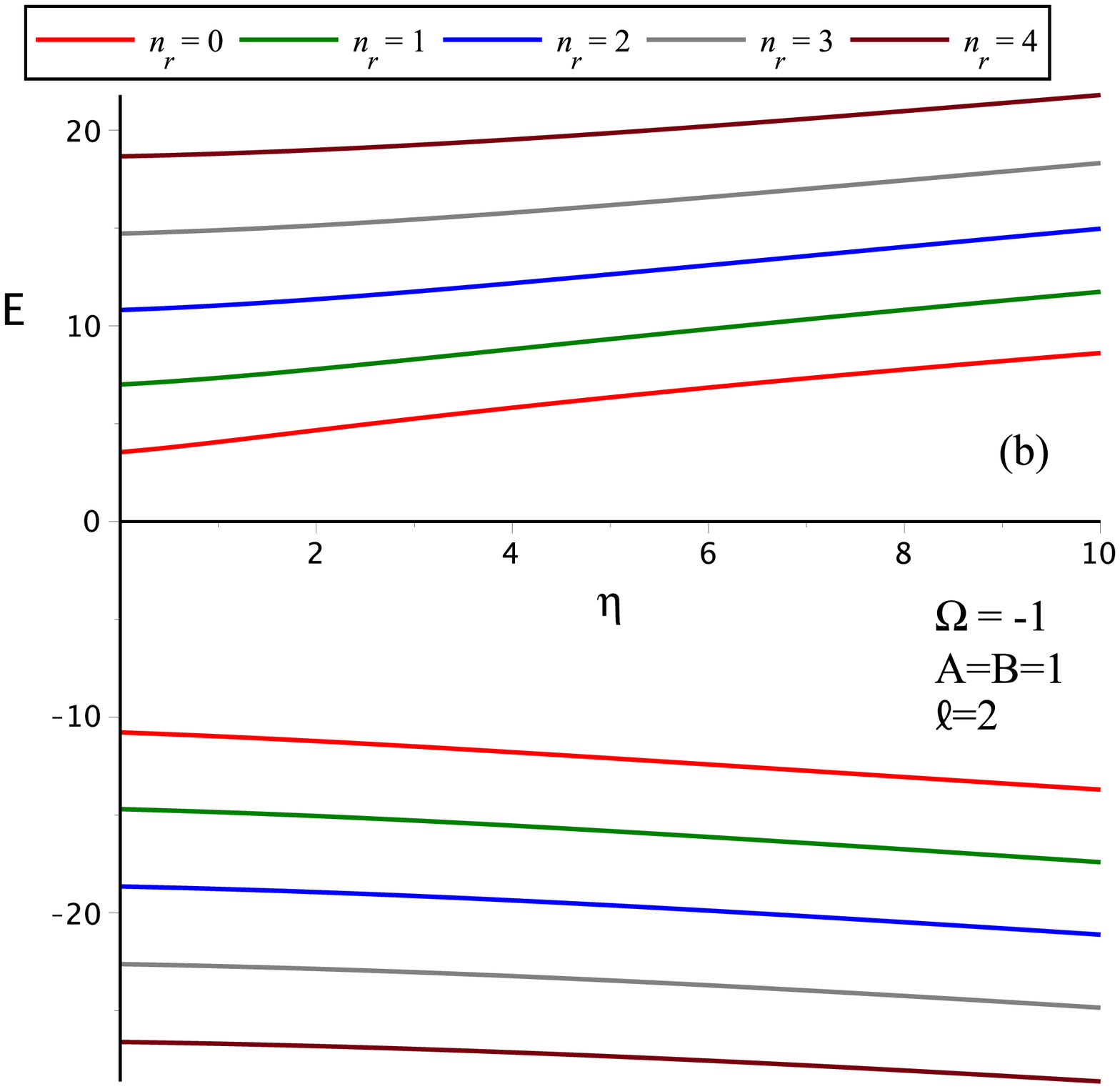}
\includegraphics[width=0.3\textwidth]{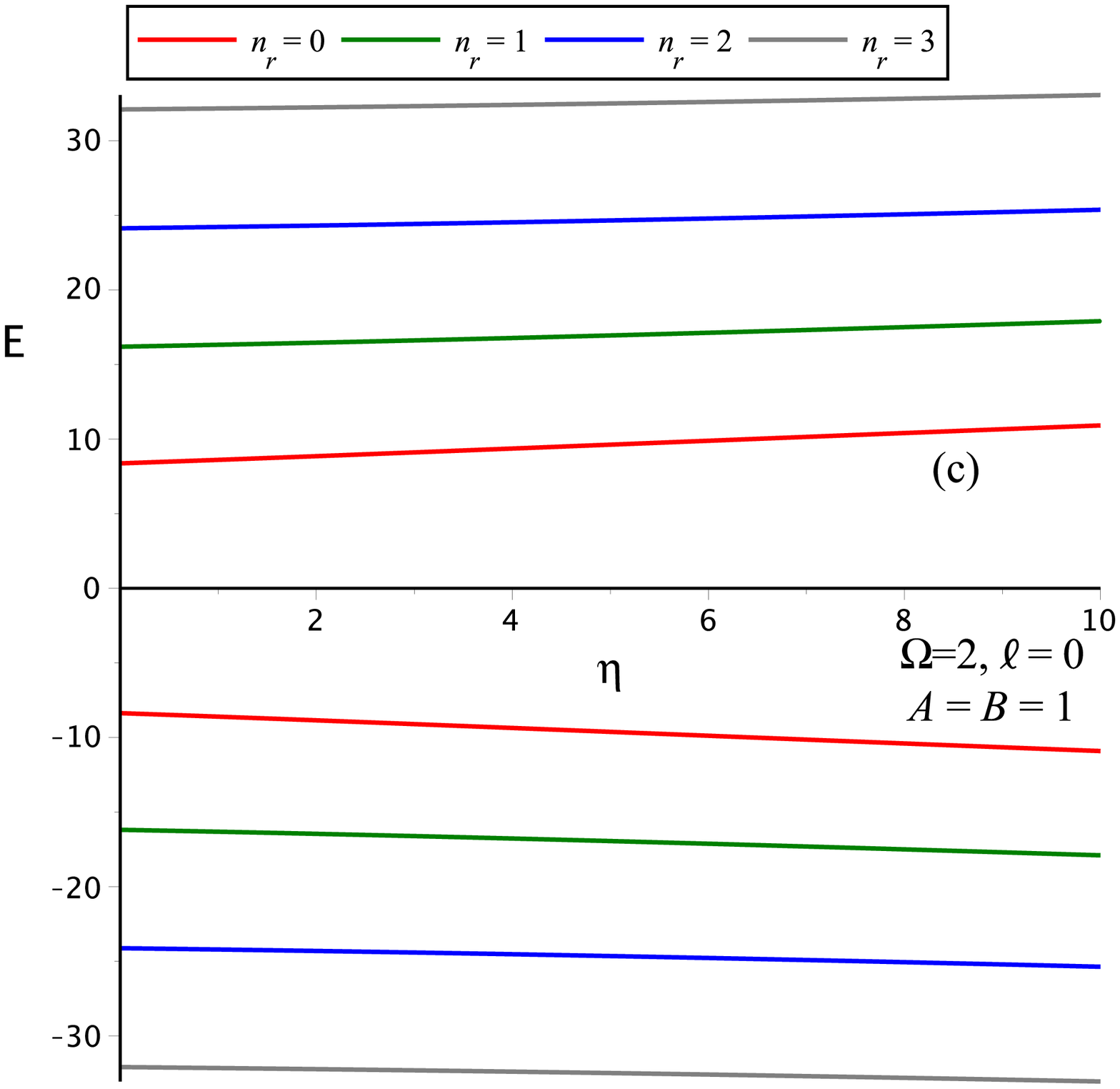}
\caption{\small 
{ For $A=B=m=1$ we show the confined KG-oscillator energies $E_{n_r,\ell}$ of (\ref{energy-eq}) against the oscillator frequency $\eta$ at $n_r=0,1,2,3,4$. In (a) for $(\Omega=1, \ell=2)$, (b) for $(\Omega=-1, \ell=2)$, and  (c) for $(\Omega=2, \ell=0)$.}}
\label{fig3}
\end{figure}%

Under such settings, one may, in a straightforward manner, obtain%
\begin{equation}
\frac{d^{2}}{d\tilde{r}^{2}}R\left( \tilde{r}\right) +\left[ \tilde{\lambda}%
-V_{eff}\left( \tilde{r}\right) \right] R\left( \tilde{r}\right) =0\,;\,V_{eff}\left( 
\tilde{r}\right) =\frac{\left( \tilde{\ell}^{2}-1/4\right) }{\tilde{r}^{2}}%
+\beta ^{2}\tilde{r}^{2}+2mA\tilde{r}+\frac{2mB}{\tilde{r}},
\label{R(rt)-eq}
\end{equation}%
which immediately inherits the form of (\ref{S(r)-solution}) so that%
\begin{eqnarray}
R\left( \tilde{r}\right) &=&R\left( \sqrt{Q\left( r\right) }r\right) =%
\mathcal{N\,}\left( \sqrt{Q\left( r\right) }r\right) ^{\left\vert \tilde{\ell%
}\right\vert +1/2}\,\exp \left( -\frac{\beta ^{2}Q\left( r\right) r^{2}+2Am%
\sqrt{Q\left( r\right) }r}{2\,\beta }\right)  \notag \\
&&\times \,H_B\left( 2\,\left\vert \tilde{\ell}\right\vert ,\frac{2mA}{%
\beta ^{3/2}},\frac{A^{2}m^{2}+\lambda \beta ^{2}}{\beta ^{3}},\frac{4mB}{%
\sqrt{\beta }},\sqrt{\beta }\sqrt{Q\left( r\right) }r\right)  \label{R(rt)}
\end{eqnarray}%
Next, with $R\left( \tilde{r}\right) =R\left( \tilde{r}\left( r\right)
\right) =m\left( r\right) ^{-1/4}\phi \left( r\right) $, equation (\ref{R(rt)-eq}) reads 
\begin{equation}
\left( \frac{1}{\sqrt{m\left( r\right) }}\frac{d}{dr}\frac{1}{\sqrt{m\left(
r\right) }}\frac{d}{dr}\right) m\left( r\right) ^{-1/4}\phi \left( r\right) +%
\left[ \lambda -V_{eff}\left( \tilde{r}\right) \right] m\left( r\right)
^{-1/4}\phi \left( r\right) =0.  \label{phi(r)-eq}
\end{equation}%
Now multiplying from the left by $m\left( r\right) ^{1/4}$ we get%
\begin{equation}
\left( m\left( r\right) ^{-1/4}\frac{d}{dr}m\left( r\right) ^{-1/2}\frac{d}{%
dr}m\left( r\right) ^{-1/4}\right) \phi \left( r\right) +\left[ \lambda
-V_{eff}\left( \tilde{r}\left( r\right) \right) \right] \phi \left( r\right) =0,
\label{MM-PDM eq}
\end{equation}%
where%
\begin{equation}
V_{eff}\left( \tilde{r}\left( r\right) \right) =\frac{\left( \tilde{\ell}%
^{2}-1/4\right) }{Q\left( r\right) r^{2}}+\beta ^{2}Q\left( r\right)
r^{2}+2mA\sqrt{Q\left( r\right) }r+\frac{2mB}{\sqrt{Q\left( r\right) }r}
\label{PDM-effective potential}
\end{equation}%
that admits isospectrality with the one dimensional Schr\"{o}dinger-like
confined KG-oscillator in (\ref{R(r)-eq-S(r)}) with its eigenvalues given in (\ref{energy-eq}) and its eigen functions given by (\ref{R(rt)}). Hence the radial wave function is eventually given by 
\begin{equation}
\psi \left( \tilde{r}\right) =\psi \left( \sqrt{Q\left( r\right) }r\right) =%
\frac{R\left( \sqrt{Q\left( r\right) }r\right) }{\sqrt{\sqrt{Q\left(
r\right) }r}}=\frac{f\left( r\right) ^{-1/4}\phi \left( r\right) }{\sqrt{%
\sqrt{Q\left( r\right) }r}}.  \label{wve function connections}
\end{equation}%
Obviously, the confined KG-oscillators of (\ref{MM-PDM eq}) and that of (\ref{R(r)-eq-S(r)}) are isospectral and share the same energy eigenvalues of (\ref{energy-eq}), therefore. Yet we may safely conclude that the two confined
KG-oscillators in a the (2+1)-dimensional G\"{u}rses to deformed-G\"{u}rses space-time backgrounds, (\ref{KG-oscillator}) and (\ref{PDM-KG1}), respectively, are invariant and isospectral.
Moreover,  one should notice that equation (\ref{MM-PDM eq}) resembles an effective
position-dependent mass (PDM) particles in the one-dimensional von Roos PDM
Hamiltonian \cite{von Roos} with Mustafa and Mazharimusavi parametric
settings \cite{Mustafa Habib 2007,Mustafa 2020,Mustafa Algadhi 2019}).%

\section{Confined-deformed KG-oscillator from a (2+1)-dimensional G\"{u}rses to a G\"{u}rses space-time backgrounds}

Let us rewrite the deformed (2+1)-dimensional G\"{u}rses space-time metric (\ref{Gurses transformed metric}), using the transformation recipe in (\ref{PT1}), as%
\begin{equation}
d\tilde{s}^{2}=-dt^{2}-2\Omega \,Q\left( r\right) r^{2}dt\,d\theta +Q\left(
r\right) r^{2}\left( 1-\Omega ^{2}\,Q\left( r\right) r^{2}\right) d\theta
^{2}+m\left( r\right) dr^{2},  \label{Gurses metric1}
\end{equation}%
and compare it with (\ref{Gurses metric0}) of \cite{Gurses 1994} to imply,
that $a_{_{0}}=1,$%
\begin{equation}
\psi =\frac{1}{m\left( r\right) },\,q=\frac{\mu }{3}Q\left( r\right)
r^{2}=c_{_{0}}+\frac{e_{_{0}}\mu }{3}r^{2}.
\end{equation}%
Consequently, our $Q(r)$ reads%
\begin{equation}
Q\left( r\right) =e_{_{0}}+\frac{3c_{_{0}}}{\mu r^{2}}\Longleftrightarrow
Q\left( r\right) r^{2}=e_{_{0}}r^{2}+\frac{3c_{_{0}}}{\mu },
\label{Q(r) Gurses}
\end{equation}%
to imply, through (\ref{m(r)-Q(r)}), that%
\begin{equation}
m\left( r\right) =\frac{\mu e_{_{0}}^{2}r^{2}}{\mu e_{_{0}}r^{2}+3c_{_{0}}}%
\Longleftrightarrow \psi =\frac{1}{m(r)}=\frac{1}{e_{_{0}}}+\frac{3c_{_{0}}}{\mu
e_{_{0}}^{2}r^{2}}.  \label{m(r) Gurses}
\end{equation}%
Obviously, the resulted parametric structures agree with those of G\"{u}rses 
\cite{Gurses 1994} in (\ref{Gurses metric0}), where%
\begin{equation}
b_{_{0}}=\frac{1}{e_{_{0}}},\,b_{_{1}}=\frac{3c_{_{0}}}{\mu e_{_{0}}^{2}},%
\text{\ }\lambda _{_{0}}=0,\text{ }h=e_{_{0}}r. \label{Gurses parameters2}
\end{equation}%
Then the corresponding metric tensor is given by%
\begin{equation}
\tilde{g}_{\mu \nu }=\left( 
\begin{tabular}{ccc}
$-1\,$ & $0$ & $-\Omega \,\left( e_{_{0}}r^{2}+\frac{3c_{_{0}}}{\mu }%
\smallskip \right) $ \\ 
$0$ & $\,\left( \frac{\mu e_{_{0}}^{2}r^{2}}{\mu e_{_{0}}r^{2}+3c_{_{0}}}%
\right) \smallskip $ & $0$ \\ 
$-\Omega \,\left( e_{_{0}}r^{2}+\frac{3c_{_{0}}}{\mu }\smallskip \right) $ & 
$0$ & $\,\,\left( e_{_{0}}r^{2}+\frac{3c_{_{0}}}{\mu }\smallskip \right)
\left( 1-\Omega ^{2}\left( e_{_{0}}r^{2}+\frac{3c_{_{0}}}{\mu }\smallskip
\right) \right) $%
\end{tabular}%
\right) ,  \label{Gurses-to-Gurses metric tensor}
\end{equation}%
with $\det (\tilde{g})=-e_{_{0}}^{2}r^{2}$, and its inverse%
\begin{equation}
\tilde{g}^{\mu \nu }=\left( 
\begin{tabular}{ccc}
$-\left( 1-\Omega ^{2}\left( e_{_{0}}r^{2}+\frac{3c_{_{0}}}{\mu }\smallskip
\right) \right) \smallskip $ & $0$ & $-\Omega \,$ \\ 
$0\smallskip $ & $\left( \frac{\mu e_{_{0}}r^{2}+3c_{_{0}}}{\mu
e_{_{0}}^{2}r^{2}}\right) \,$ & $0$ \\ 
$-\Omega \,$ & $0$ & $\,\,\frac{1}{\left( e_{_{0}}r^{2}+\frac{3c_{_{0}}}{\mu 
}\smallskip \right) }$%
\end{tabular}%
\right)  \label{Gurses-to-Gurses inverse metric tensor}
\end{equation}%
Moreover, we may now report the corresponding confined-deformed KG-oscillator in the deformed G\"{u}rses space-time (\ref{Gurses metric1}) background as%
\begin{equation}
\left( \left( \frac{\mu e_{_{0}}^{2}r^{2}}{\mu e_{_{0}}r^{2}+3c_{_{0}}}%
\right) ^{-1/4}\frac{d}{dr}\left( \frac{\mu e_{_{0}}^{2}r^{2}}{\mu
e_{_{0}}r^{2}+3c_{_{0}}}\right) ^{-1/2}\frac{d}{dr}\left( \frac{\mu
e_{_{0}}^{2}r^{2}}{\mu e_{_{0}}r^{2}+3c_{_{0}}}\right) ^{-1/4}\right) \phi
\left( r\right) +\left[ \tilde{\lambda}-V\left( \tilde{r}\left( r\right)
\right) \right] \phi \left( r\right) =0,  \label{KGO1}
\end{equation}%
where%
\begin{equation}
V\left( r\right) =V\left( \tilde{r}\left( r\right) \right) =\frac{\left( 
\tilde{\ell}^{2}-1/4\right) }{\left( e_{_{0}}+\frac{3c_{_{0}}}{\mu r^{2}}%
\right) r^{2}}+\beta ^{2}\left( e_{_{0}}+\frac{3c_{_{0}}}{\mu r^{2}}\right)
r^{2}+2mA\sqrt{\left( e_{_{0}}+\frac{3c_{_{0}}}{\mu r^{2}}\right) }r+\frac{%
2mB}{\sqrt{\left( e_{_{0}}+\frac{3c_{_{0}}}{\mu r^{2}}\right) }r}.
\label{V1(r)-KGO}
\end{equation}%
This confined-deformed KG-oscillator is isospectral and invariant with that of (\ref{R(r)-eq-S(r)})
and, therefore, shares the eigen energies given in (\ref{energy-eq}). The
corresponding eigenfunctions are readily given in (\ref{R(rt)}) with $%
Q\left( r\right) $ and $m\left( r\right) $ are as defined in (\ref{Q(r) Gurses}) and (\ref{m(r) Gurses}), respectively. The confined KG-oscillators, (\ref{KGO1}) and (\ref{R(r)-eq-S(r)}), in G\"{u}rses space-time backgrounds are invariant and isospectral, therefore.

\section{Confined-deformed KG-oscillator from a pseudo-G\"{u}rses to a G\"{u}rses space-time backgrounds}

In this section, we consider a deformation in the radial coordinate through%
\begin{equation}
\sqrt{Q\left( r\right) }r=ar+\frac{b}{r}\Longleftrightarrow Q\left( r\right)
r^2=\left( ar+\frac{b}{r}\right) ^{2},  \label{Q(r)2}
\end{equation}%
which would, through the correlation in (\ref{m(r)-Q(r)}), imply that%
\begin{equation}
m\left( r\right) =\left( a-\frac{b}{r^{2}}\right) ^{2}.  \label{m(r)2}
\end{equation}%
Consequently, equation (\ref{MM-PDM eq}) yields%
\begin{equation}
\left( \left( a-\frac{b}{r^{2}}\right) ^{-1/2}\frac{d}{dr}\left( a-\frac{b}{%
r^{2}}\right) ^{-1}\frac{d}{dr}\left( a-\frac{b}{r^{2}}\right)
^{-1/2}\right) \phi \left( r\right) +\left[ \lambda -V_{1}\left( r\right) %
\right] \phi \left( r\right) =0,  \label{KG-PDEM1}
\end{equation}%
where $V\left( r\right) $ is now given by (\ref{PDM-effective potential}) as 
\begin{equation}
V_{1}\left( r\right) =\frac{\left( \tilde{\ell}^{2}-1/4\right) }{\left(
ar+b/r\right) ^{2}}+\beta ^{2}\left( ar+\frac{b}{r}\right) ^{2}+2mA\left( ar+%
\frac{b}{r}\right) .  \label{V1(r)}
\end{equation}%
Such confined-deformed KG-oscillators, (\ref{KG-PDEM1}) and (\ref{R(r)-eq-S(r)}),
in pseudo-G\"{u}rses to G\"{u}rses space-time backgrounds are invariant and
isospectral, therefore. Moreover, this system corresponds to a
(2+1)-dimensional pseudo-G\"{u}rses deformed space-time metric (\ref{Gurses transformed metric}) of the form%
\begin{equation}
d\tilde{s}^{2}=-\left( dt+\Omega \,\left( ar+\frac{b}{r}\right) ^{2}d\theta
\right) ^{2}+\left( ar+\frac{b}{r}\right) ^{2}d\theta ^{2}+\left( a-\frac{b}{%
r^{2}}\right) ^{2}dr^{2},  \label{Gurses metric 1}
\end{equation}%
The notion of pseudo-G\"{u}rses space-time metric is manifested by the fact
that this metric may yield a G\"{u}rses space-time like metric (\ref{Gurses metric}) (discussed in section 1) if the transformation is reversed. Having said that, it is obvious that the only feasible G\"{u}rses to G\"{u}%
rses space-time metric backgrounds case is the one discussed section 4,
where $Q\left( r\right) $ and $m\left( r\right) $ are, respectively, given
by (\ref{Q(r) Gurses}) and (\ref{m(r) Gurses}). Any other structure for $%
Q\left( r\right) $ and $m\left( r\right) $ in (\ref{Gurses transformed
metric}) should be classified as pseudo-G\"{u}rses to G\"{u}rses space-time
metric backgrounds.
\begin{figure}[!ht]  
\centering
\includegraphics[width=0.3\textwidth]{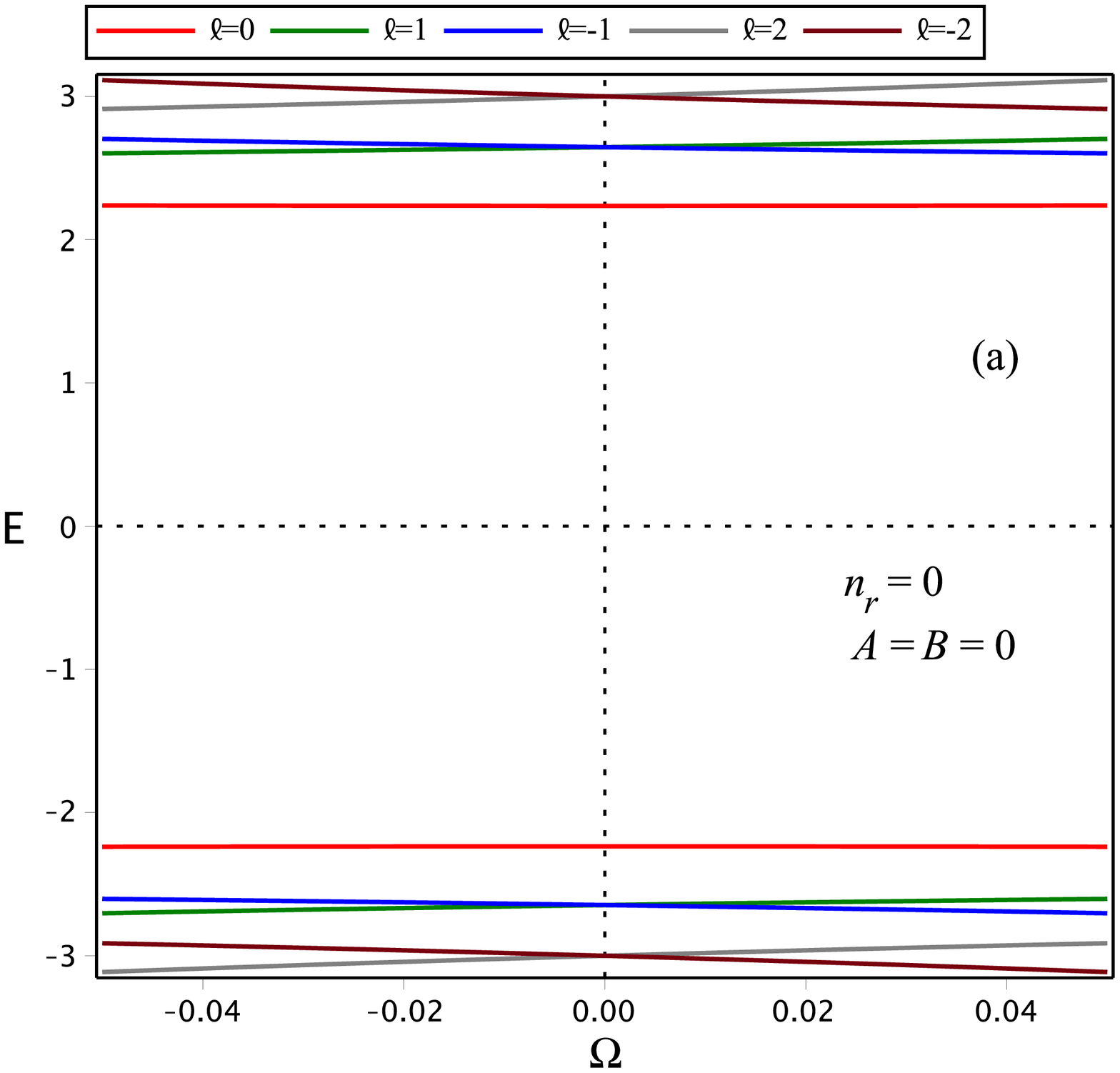}
\includegraphics[width=0.3\textwidth]{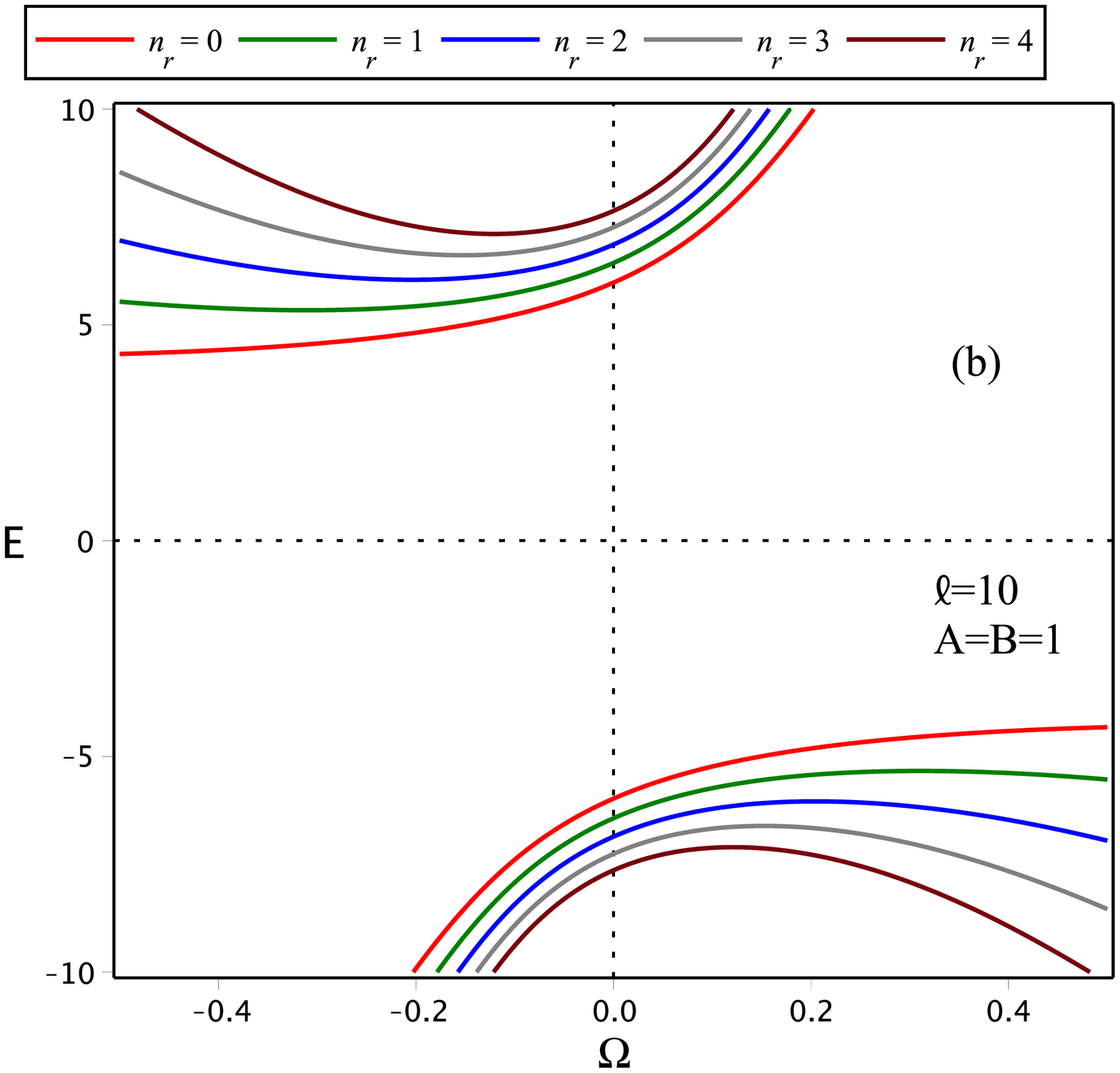}
\includegraphics[width=0.3\textwidth]{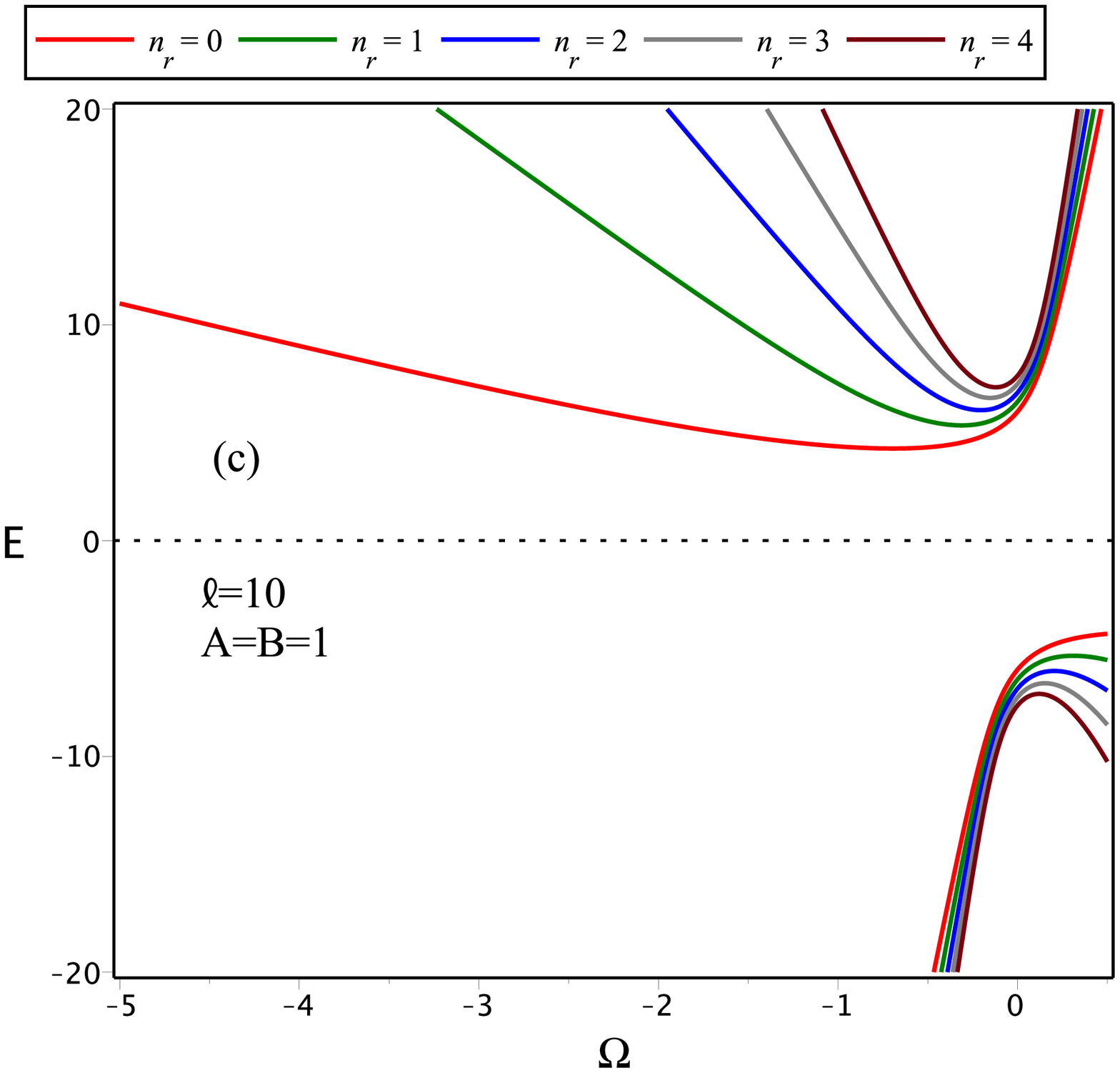}
\caption{\small 
{ (a) corresponds to Fig.1(a) but with the vorticity parameter $\Omega\in [-0.05,0.05]$, (b) corresponds to Fig.3(e) but with the vorticity parameter $\Omega\in [-0.5,0.5]$, and (c) corresponds to 4(b) above but with $\Omega\in [-5,0.5]$.}}
\label{fig4}
\end{figure}%

\section{Concluding remarks}
In this study, we have considered (in section 2) the KG-oscillator confined in a Cornell-type scalar potential $S\left( r\right) $ of (\ref{Cornell potential}) in some (2+1)-dimensional G\"{u}rses space-time backgrounds. We started with a confined KG-oscillator in a G\"{u}rses space-time background (i.e. G\"{u}rses space-time metric $ds^{2}$ (\ref{Gurses metric}) at specific parametric settings) and reported the corresponding exact solution in (\ref{S(r)-solution}) and (\ref{energy-eq}). Hereby, we argue that the natural tendency of a physically viable solution of a more general case ( confined KG-oscillator in our case) should necessarily collapse into that of a less complicated KG-oscillator ones in (\ref{HO-lambda}) and (\ref{HO-R(r)}), when the confinement parameters are switched off. So is the tendency of our reported exact solution in (\ref{S(r)-solution}) and (\ref{energy-eq}). 

To observe the effect of the vorticity parameter $\Omega=-\mu/3$ (in our case) on the energy levels  $E_{n_r,\ell}$, we have reported $E_{n_r,\ell}$ (\ref{energy-eq}) versus $\Omega$ in Figures 1-2, and $E_{n_r,\ell}$ vs $\eta$ (the KG-oscillator frequency) in Figure 3. The mathematical structure as well as the reported figures document the critical role of the second term, $(-2\,\Omega \,\ell \,E)$ on the L.H.S. of (\ref{energy-eq}), in shaping the energies of the confined KG-oscillator. This effect is summarized as follows: (i) the positive energy $E_+$, with $\Omega$ and $\ell$ are both positive or both negative, is boosted upward for $\ell\neq0$, (ii) the negative energy  $E_-$, with positive/negative $\Omega$ and negative/positive $\ell$, respectively, is boosted downward for $\ell\neq0$, and (iii) for a fixed $\Omega$ the energy gap is shifted upwards (for positive $\Omega$ and $\ell$) and downwards (for negative $\Omega$ and $\ell$), whilst the gap increases as the KG-oscillator frequency $\eta$ increases. As a result of such effects,  the energy levels crossings  (documented in Fig.1), the energy levels clustering (documented in Fig.2(b)-2(f)), and the energy gap shifting (documented in Fig.3(a) and 3(b)) are unavoidable natural manifestations in the process. Moreover, to figure out how the energy levels behave near $\Omega=0$ (i.e., flat space-time background) we have also reported Figure 4. Figure 4(a) corresponds to Fig.1(a) but with the vorticity parameter $\Omega\in [-0.05,0.05]$. Obviously, at $\Omega=0$ the energy levels admit degeneracy as follows: $E_{0,0},E_{0,\pm1},E_{0,\pm2}$ form bottom to top for the positive energies and from top to bottom for negative energies. As we move far from $\Omega=0$, we see that $E_{0,+1}$ and $E_{0,-1}$ switch their ordering in moving from $\Omega>0$ to $\Omega<0$ regions (similar behaviour occurs for all $E_{n_{r}>0,+|\ell|}$ and $E_{n_{r}>0,-|\ell|}$ states). To make sure that energy levels clustering do not indulge levels crossings we plot Fig.s 4(b) and 4(c), they correspond to Fig.3(e) but with the vorticity parameter $\Omega\in [-0.5,0.5]$ and  $\Omega\in [-5,0.5]$, respectively. Clearly, for a fixed of the magnetic quantum number $\ell=\pm|\ell|$, no energy levels crossings are observed and the energy levels just cluster without ordering change.

Next, we have considered (in section 3) the confined KG-oscillator in a general deformation of the (2+1)-dimensional G\"{u}rses space-time background (\ref{Gurses transformed metric}). Hereby, we have shown that the resulting confined and deformed KG-oscillator is invariant and isospectral with of the confined KG-oscillator (\ref{R(r)-eq-S(r)}) in the (2+1)-dimensional G\"{u}rses space-time background (\ref{Gurses metric}). We have further considered (in section 4) a  confined and deformed KG-oscillator in a deformed (2+1)-dimensional G\"{u}rses  (\ref{Gurses transformed metric}) to a G\"{u}rses space-time backgrounds (\ref{Gurses metric0}). That is, the deformation in the (2+1)-dimensional G\"{u}rses space-time metric $d\tilde{s}^{2}$ (\ref{Gurses transformed metric}) is chosen so that it belongs to
a another G\"{u}rses space-time metric (but with different  G\"{u}rses-type parametric setting) in (\ref{Gurses metric0}).  We have shown that such a confined and deformed KG-oscillator is invariant and isospectral with of the confined KG-oscillator (\ref{R(r)-eq-S(r)}) in the (2+1)-dimensional G\"{u}rses space-time background (\ref{Gurses metric}). Finally, we have considered (in section 5) a confined and deformed KG-oscillator in pseudo-G\"{u}rses to G\"{u}rses space-time backgrounds. The pseudo-G\"{u}rses space-time metric is manifested by the
fact that its parametric settings do not belong to G\"{u}rses space-time (\ref{Gurses metric0}) metric but it can be transformed into G\"{u}rses space-time metric (\ref{Gurses metric}) within the transformation (\ref{PT1}). Once again we have shown that such a confined and deformed KG-oscillator is invariant and isospectral with of the confined KG-oscillator (\ref{R(r)-eq-S(r)}) in the (2+1)-dimensional G\"{u}rses space-time background (\ref{Gurses metric}).

To the best of our knowledge, within the above methodical proposal settings, such KG-oscillators in the backgrounds of (2+1)-dimensional G\"{u}rses to  G\"{u}rses or to pseudo-G\"{u}rses  space-time metric have never been reported elsewhere.

\end{document}